\documentclass[twocolumn]{aastex631}

\newcommand{\todo}[1]{\textcolor{black}{#1}}
\newcommand{\inprep}[1]{\textcolor{black}{#1}} 
\newcommand{\glswithcite}[2]{%
    \ifglsused{#1}{\gls{#1} (#2)}{\glsentrylong{#1} (\glsentryshort{#1}, #2)\glsunset{#1}} 
}

\usepackage[acronym,nohypertypes={acronym}]{glossaries} 
\newacronym{LSST}{LSST}{Legacy Survey of Space and Time}
\newacronym{HST}{HST}{Hubble Space Telescope}
\newacronym{CEERS}{CEERS}{Cosmic Evolution Early Release Science Survey}
\newacronym{GZ}{GZ}{Galaxy Zoo}
\newacronym{NIR}{NIR}{near-infrared}
\newacronym{IR}{IR}{infrared}
\newacronym{NIRCam}{NIRCam}{Near-Infrared Camera}
\newacronym{DESI}{DESI}{Dark Energy Spectroscopic Instrument}
\newacronym{TIMER}{TIMER}{Time Inference with MUSE in Extragalactic Rings}
\newacronym{JADES}{JADES}{JWST Advanced Deep Extragalactic Survey}
\newacronym{DECaLS}{DECaLS}{Dark Energy Camera Legacy Survey}
\newacronym{FWHM}{FWHM}{full width at half maximum}
\newacronym{PSF}{PSF}{point spread function}

\shorttitle{Bar fractions up to $z \sim 4.0$}
\shortauthors{G\'eron et al.}
\graphicspath{{./}{figures/}}

\begin{document}

\title{Galaxy Zoo CEERS: Bar fractions up to $z \sim 4.0$}

\correspondingauthor{Tobias G\'eron}
\email{tobias.geron@utoronto.ca}

\author[0000-0002-6851-9613]{Tobias G\'eron}
\affiliation{Dunlap Institute for Astronomy \& Astrophysics, University of Toronto, 50 St. George Street, Toronto, ON M5S 3H4, Canada}

\author[0000-0001-6417-7196]{R. J. Smethurst}
\affiliation{Oxford Astrophysics, Department of Physics, University of Oxford, Denys Wilkinson Building, Keble Road, Oxford, OX1 3RH, UK}

\author[0000-0003-0475-008X]{Hugh Dickinson}
\affiliation{School of Physical Sciences, The Open University, Milton Keynes, MK7 6AA, UK}

\author[0000-0002-1067-8558]{L. F. Fortson}
\affiliation{School of Physics and Astronomy, University of Minnesota, Minneapolis, Minnesota, 55455, USA}

\author[0000-0002-3887-6433]{Izzy L. Garland}
\affiliation{Department of Theoretical Physics and Astrophysics, Faculty of Science, Masaryk University, Kotl\'{a}\v{r}sk\'{a} 2, Brno, 611 37, Czech Republic}

\author[0000-0001-8010-8879]{Sandor Kruk}
\affiliation{European Space Agency (ESA), European Space Astronomy Centre (ESAC), Camino Bajo del Castillo s/n, 28692, Villaneuva de la Cañada, Madrid, Spain}

\author[0000-0001-5578-359X]{Chris Lintott}
\affiliation{Oxford Astrophysics, Department of Physics, University of Oxford, Denys Wilkinson Building, Keble Road, Oxford, OX1 3RH, UK}

\author[0009-0009-6545-8710]{Jason Shingirai Makechemu}
\affiliation{Department of Physics, Lancaster University, Lancaster, LA1 4YB, UK}

\author[0000-0002-6016-300X]{Kameswara Bharadwaj Mantha}
\affiliation{School of Physics and Astronomy, University of Minnesota, Minneapolis, Minnesota, 55455, USA}
\affiliation{Minnesota Institute for Astrophysics, University of Minnesota, Minneapolis, Minnesota, 55455, USA}

\author[0000-0003-0846-9578]{Karen L. Masters}
\affiliation{Department of Physics and Astronomy, Haverford College, 370 Lancaster Avenue, Ardmore, Pennsylvania 19041, USA}

\author[0000-0003-1217-4617]{David O'Ryan}
\affiliation{Centro de Astrobiología, INTA-CSIC, Camino Bajo del Castillo, s/n, 28692, Villanueva de la Cañada, Madrid, Spain}

\author[0000-0003-0046-9848]{Hayley Roberts}
\affiliation{School of Physics and Astronomy, University of Minnesota, Minneapolis, Minnesota, 55455, USA}
\affiliation{Minnesota Institute for Astrophysics, University of Minnesota, Minneapolis, Minnesota, 55455, USA}

\author[0000-0001-5882-3323]{B.D. Simmons}
\affiliation{Department of Physics, Lancaster University, Lancaster, LA1 4YB, UK}

\author[0000-0002-6408-4181]{Mike Walmsley}
\affiliation{Dunlap Institute for Astronomy \& Astrophysics, University of Toronto, 50 St. George Street, Toronto, ON M5S 3H4, Canada}

\author[0000-0003-2536-1614]{Antonello Calabr\`o}
\affiliation{INAF Osservatorio Astronomico di Roma, Via Frascati 33, 00078 Monte Porzio Catone, Rome, Italy}

\author[0000-0002-3445-855X]{Rimpei Chiba}
\affiliation{Canadian Institute for Theoretical Astrophysics, University of Toronto, 60 St. George Street, Toronto, ON M5S 3H8, Canada}

\author[0000-0001-6820-0015]{Luca Costantin}
\affiliation{Centro de Astrobiología (CAB), CSIC-INTA, Ctra. de Ajalvir km 4, Torrejón de Ardoz, E-28850, Madrid, Spain}

\author[0000-0001-7081-0082]{Maria R. Drout}
\affiliation{David A. Dunlap Department of Astronomy \& Astrophysics, University of Toronto, 50 St. George Street, Toronto, ON, M5S 3H4, Canada}

\author[0000-0002-0897-3013]{Francesca Fragkoudi}
\affiliation{Institute for Computational Cosmology, Department of Physics, Durham University, South Road, Durham, DH1 3LE, UK}

\author[0000-0002-4162-6523]{Yuchen Guo}
\affiliation{Department of Astronomy, The University of Texas at Austin, Austin, TX, USA}

\author[0000-0002-4884-6756]{B. W. Holwerda}
\affiliation{Department of Physics and Astronomy, University of Louisville, Natural Science Building 102, 40292 KY Louisville, USA}

\author[0000-0002-1590-0568]{Shardha Jogee}
\affiliation{Department of Astronomy, The University of Texas at Austin, Austin, TX, USA}

\author[0000-0002-6610-2048]{Anton M. Koekemoer}
\affiliation{Space Telescope Science Institute, 3700 San Martin Drive, Baltimore, MD 21218, USA}

\author[0000-0003-1581-7825]{Ray A. Lucas}
\affiliation{Space Telescope Science Institute, 3700 San Martin Drive, Baltimore, MD 21218, USA}

\author[0000-0001-9879-7780]{Fabio Pacucci}
\affiliation{Center for Astrophysics $\vert$ Harvard \& Smithsonian, 60 Garden St, Cambridge, MA 02138, USA}
\affiliation{Black Hole Initiative, Harvard University, 20 Garden St, Cambridge, MA 02138, USA}



\begin{abstract}

  We study the evolution of the bar fraction in disc galaxies between $0.5 < z < 4.0$ using multi-band coloured images from JWST CEERS. These images were classified by citizen scientists in a new phase of the Galaxy Zoo project called GZ CEERS. Citizen scientists were asked whether a strong or weak bar was visible in the host galaxy. After considering multiple corrections for observational biases, we find that the bar fraction decreases with redshift in our volume-limited sample (n = \todo{398}); from \todo{$25^{+6}_{-4}$\%} at $0.5 < z < 1.0$ to \todo{$3^{+6}_{-1}$\%} at $3.0 < z < 4.0$. However, we argue it is appropriate to interpret these fractions as lower limits. Disentangling real changes in the bar fraction from detection biases remains challenging. Nevertheless, we find a significant number of bars up to $z = 2.5$. This implies that discs are dynamically cool or baryon-dominated, enabling them to host bars. This also suggests that bar-driven secular evolution likely plays an important role at higher redshifts. When we distinguish between strong and weak bars, we find that the weak bar fraction decreases with increasing redshift. In contrast, the strong bar fraction is constant between $0.5 < z < 2.5$. This implies that the strong bars found in this work are robust long-lived structures, unless the rate of bar destruction is similar to the rate of bar formation. Finally, our results are consistent with disc instabilities being the dominant mode of bar formation at lower redshifts, while bar formation through interactions and mergers is more common at higher redshifts.

\end{abstract}











\section{Introduction}

Bars are common and important structures in galaxies that are able to influence the evolution of their hosts in multiple ways. They transfer angular momentum from the inner disc to the outer disc and dark matter halo \citep{lynden_bell_1972,sellwood_1981,athanassoula_2003,athanassoula_2013}, while funneling gas from the outskirts of the galaxy along the arms of the bar to the centre of the galaxy \citep{sorensen_1976,athanassoula_1992,davoust_2004,villa-vargas_2010,fragkoudi_2016,vera_2016,spinoso_2017,george_2019}. This inflow of gas could cause a central starburst \citep{jogee_2005,sheth_2005,hunt_2008} and potentially trigger an AGN \citep{fanali_2015,galloway_2015,garland_2024}. Bars have been linked to quenching, as they appear more often in quiescent galaxies (i.e. red, massive and gas-poor galaxies; \citealt{hoyle_2011,masters_2011, masters_2012,cheung_2013,vera_2016,cervantessodi_2017,kruk_2018,fraser_mckelvie_2020b}). Many studies allow for the possibility that bars are robust structures that are long-lived \citep{jogee_2004,debattista_2006,athanassoula_2013,gadotti_2015, perez_2017,ghosh_2023,fragkoudi_2025,lopez_2024}, although some suggest that they might be short-lived transient features and can get destroyed or weakened \citep{bournaud_2002,shen_2004,athanassoula_2005b, bournaud_2005,ghosh_2021}. In summary, bars are potentially long-lived structures that affect their host galaxies in multiple significant ways. Having a clear understanding of the prevalence of bars at different redshifts is crucial to understand how long bar-driven effects have been important in galaxy evolution.

There are many different kinds of bars. \citet{devaucouleurs_1959, devaucouleurs_1963} noted that some bars are obvious and long, which he termed `strong bars'. In contrast, other bars were faint and small, which he termed `weak bars'. This idea of bar strength can be measured in many different ways. For example, as stronger bars have more elongated isophotes, the maximum ellipticity of these isophotes has been used to estimate bar strength \citep{athanassoula_1992b,laurikainen_2002,erwin_2004}. In a previous iteration of Galaxy Zoo (GZ) \citep{walmsley_2022}, citizen scientists were asked to judge bar strength and classify whether a galaxy had a strong or weak bar using images from the Dark Energy Camera Legacy Survey (DECaLS, \citealp{dey_2019}). Volunteers needed to complete a brief tutorial before being able to classify galaxies. In addition, volunteers had access to a `field guide' that shows examples of strongly and weakly barred galaxies. These classifications were used to classify bars as `strong' or `weak'.

Bars are common structures among disc galaxies. The combined weak and strong bar fraction in low-redshift studies using optical wavelengths is around 43\%-52\% \citep{marinova_2007,barazza_2008,aguerri_2009,buta_2019,geron_2021}. This fraction rises to 59-73\% when using infrared wavelengths \citep{eskridge_2000,marinova_2007,menendez_delmestre_2007,sheth_2008}, possibly because these wavelengths are less affected by star formation and dust \citep{erwin_2018}. Lower bar fractions of 23.6 - 29.4\% are found when only considering strong bars \citep{masters_2011, skibba_2012, cheung_2013}. 

Studying the bar fraction at higher redshifts is challenging. It has been proposed that the so-called `epoch of bar formation' occurred around $z \sim 0.7 - 1$, when galaxies started to dynamically cool and become more disc dominated, allowing them to form and maintain bars \citep{kraljic_2012, melvin_2014,simmons_2014}. This period coincides with an observed lower rate of major mergers \citep{conselice_2003,ryan_2008,jogee_2009,lotz_2011}. Mergers can destroy bars \citep{casteels_2013,guedes_2013,ghosh_2021}, although mergers and tidal interactions can also trigger bar formation \citep{noguchi_1987,elmegreen_1991,lang_2014,peschken_2019,merrow_2024}. Either way, secular processes, such as bar quenching, are predicted to become increasingly more important for the evolution of galaxies from that epoch onwards \citep{kraljic_2012,melvin_2014,simmons_2014}. This epoch has been studied in detail with the \gls{HST}. Bar fractions at these higher redshifts (0.5 $<$ $z$ $<$ 2) typically vary between 10-20\% \citep{sheth_2008,simmons_2014}. However, there is some debate about whether the bar fraction falls or remains constant over this range. For example, \citet{melvin_2014} find that the bar fraction decreases from 22\% at $z$ = 0.4 to 11\% at $z$ = 1, while \citet{elmegreen_2004} find a constant bar fraction of 23\% from $z$ = 0 to $z$ = 1.1. \citet{simmons_2014} also find that the bar fraction across 0.5 $<$ $z$ $<$ 2 does not significantly evolve.

More recently, JWST made it possible to study bars at even higher redshifts due to its higher angular resolution and sensitivity in the \gls{NIR}, which makes identifying bars at higher redshifts easier. \citet{guo_2023} have found multiple bars up to $z$ $\sim$ 2 using the JWST \glswithcite{CEERS}{\citealt{finkelstein_2023}}, indicating that bars were present $\sim$8-10 Gyr ago. More bars are being found using JWST between redshifts of 2.4 $<$ $z$ $<$ 4.2 \citep{huang_2023, costantin_2023, smail_2023, amvrosiadis_2025}. Although these studies clearly demonstrate that bars can exist at these high redshifts, they offer no insight into their prevalence. \citet{leconte_2024} addressed this problem using a sample of 339 disc galaxies in \gls{CEERS} and found that the bar fraction decreases from $18^{+5}_{-5}$\% between 1 $\leq$ $z$ $\leq$ 2 to $14^{+7}_{-6}$\% at 2 $<$ $z$ $\leq$ 3. Similarly, \citet{guo_2024} report observed bar fractions $\leq$ 10\% at $z$ $\sim$ $2-4$ in \gls{CEERS}, but they point out that the true bar fraction could be higher as their study cannot robustly detect bars with semi-major axis below 1.5 kpc. A lower bar fraction at higher redshifts is expected, as a lot of cosmological simulations predict very few bars beyond $z$ = 1 $\sim$ 1.5, because the disc is thought to be too dynamically hot to form bars \citep{kraljic_2012, reddish_2022}. Interestingly, \citet{bland_hawthorn_2023,bland_hawthorn_2024} have suggested that bars can actually form in dynamically hot galaxies, depending on ratio of baryons to dark matter of the disc. Other recent simulations also find that bars can form at higher redshifts \citep{fragkoudi_2020, fragkoudi_2025, zana_2022,rosas_guevara_2022}.

Traditionally, experts would visually inspect galaxies to describe their morphologies and find bars (e.g. see \citealt{nair_2010}). However, multiple iterations of the Galaxy Zoo project have shown that citizen scientists are also very capable of reliably describing the morphology of galaxies and identifying bars \citep{lintott_2008,willett_2013}. These volunteer classifications were used in \citet{walmsley_2022} to train an ensemble of Bayesian convolutional neural networks to automatically classify a large sample of galaxies. However, a sufficiently large sample with accurate morphological classifications needed to train such a network for JWST does not yet exist. This shows the need for a citizen science project to examine JWST images. This was done in GZ CEERS, a pilot project that looked at publicly available JWST \gls{CEERS} \gls{NIRCam} images. GZ CEERS has just finished classifying ~7000 galaxies (\inprep{Masters et al. in prep}), which we will use in this study to find bars. Other reliable ways to detect bars also exist. For example, ellipse fitting is a commonly used method \citep{jedrzejewski_1987, wozniak_1995,elmegreen_2004,jogee_2004,marinova_2007,sheth_2008}, where multiple ellipses are fitted to images of galaxies. A bar is found if the ellipticity and position angle of the ellipses change as expected when a bar is present. This method has been used successfully in other JWST studies to identify bars (e.g. see \citealt{guo_2023,guo_2024,leconte_2024,pritchett_2024}). \citet{lee_2019} show that ellipse fitting techniques miss $\sim$15\% of bars compared to visual classification at low redshifts, while \citet{guo_2024} find similar results between the two techniques at $z$ $\sim 2-4$.

In this work, we use volunteer classifications of multi-band coloured \gls{CEERS} images to identify barred galaxies. We do this for a wide redshift range (\todo{0.5} $<$ $z$ $<$ \todo{4.0}) to get an accurate understanding of how common bars are in the high redshift Universe, and whether this differs for strong and weak bars. This work will help to investigate the importance of secular processes at high redshifts and when discs become dynamically cool. It will also provide insight into the lifetime and robustness of strong and weak bars. The structure of this paper is as follows: in Section \ref{sec:data_methods}, we describe the data and methods used in this work. The results are presented in Section \ref{sec:results} and discussed in Section \ref{sec:discussion}. Finally, we summarise the main conclusions in Section \ref{sec:conclusions}. We assume a standard flat $\Lambda$CDM cosmological model with H$_{0}$ = 70 km s$^{-1}$ Mpc$^{-1}$, $\Omega_{\rm m}$ = 0.3 and $\Omega_{\rm \Lambda}$ = 0.7 where necessary, implemented with \texttt{Astropy} \citep{astropy_2013,astropy_2018,astropy_2022}.

\section{Data and methods}
\label{sec:data_methods}

We summarise GZ CEERS in Section \ref{sec:gz_jwst_ceers} and the details of our sample selection in Section \ref{sec:sample_selection}. In Section \ref{sec:bar_fraction}, we discuss how we calculate bar fractions, the corrections applied to account for biases, and how we compute the associated error bars.

\subsection{Galaxy Zoo CEERS}
\label{sec:gz_jwst_ceers}

GZ CEERS used \gls{NIRCam} observations from the \gls{CEERS} data release 0.5 \citep{bagley_2023, doi_finkelstein_2023}. This included imaging in six broadband filters (F115W, F150W, F200W, F277W, F356W and F444W) and one medium-band filter (F410M) over four pointings (1, 2, 3 and 6). The \gls{CEERS} data can be found on MAST at \dataset[10.17909/z7p0-8481]{https://archive.stsci.edu/doi/resolve/resolve.html?doi=10.17909/z7p0-8481}. The targets were selected by running SEP (Source Extractor for Python, \citealp{bertin_1996, barbary_2016}) on every pointing. To be consistent with previous iterations of Galaxy Zoo, we used coloured images instead of single-band images. This allows us to take advantage of the different properties of the different bands. The smaller \gls{FWHM} of the shorter wavelength bands is better suited to identify smaller bars, while the longer wavelength bands are less sensitive to dust extinction and star formation effects. We used the publicly released colour images created by the \gls{CEERS} team, which were generated using the seven different filters mentioned above\footnote{https://ceers.github.io/ceers-first-images-release.html}. The F115W and F150W filters are visualised in blue, the F200W and F277W in green, F356W in orange and F410M and F444W in red. See \inprep{Masters et al. (in prep)} for more information on the image processing for GZ CEERS.

\begin{figure}
	\includegraphics[width=\columnwidth]{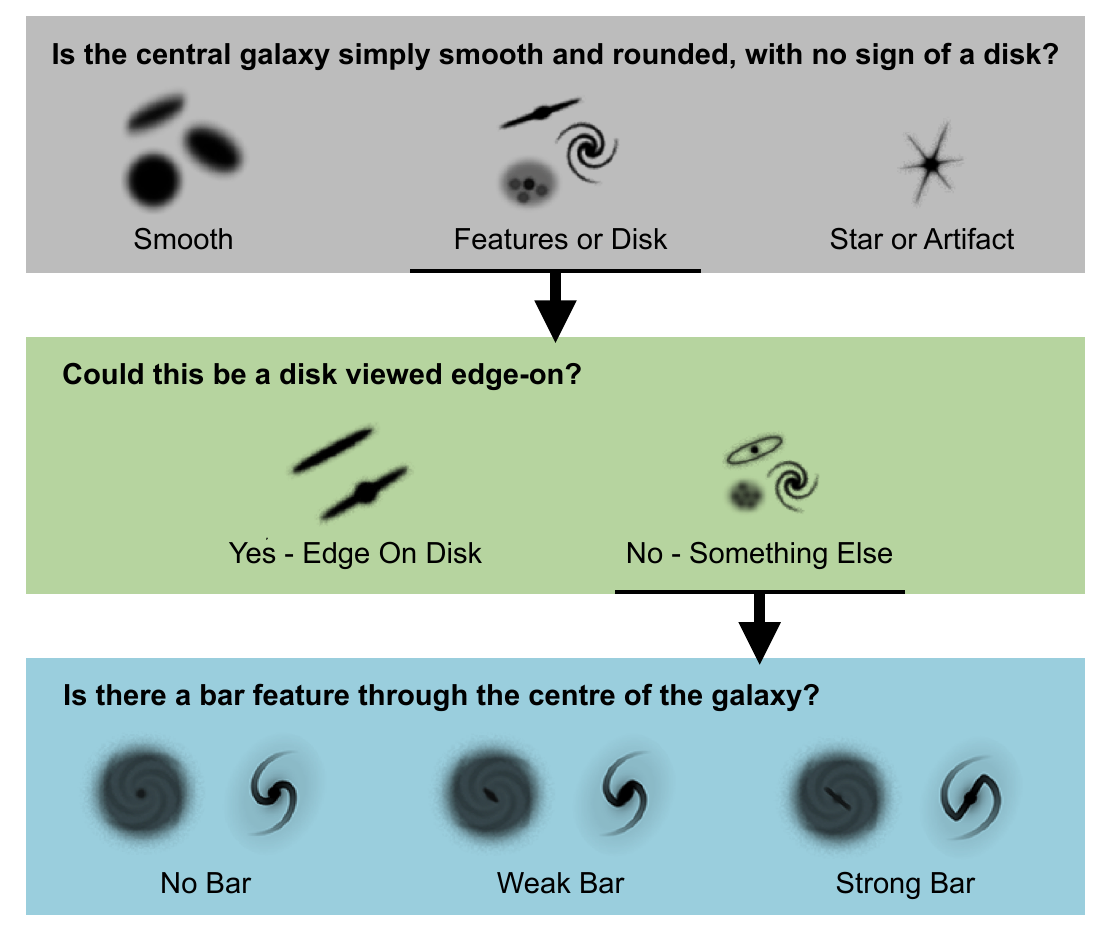}
    \caption{A subset of the decision tree used in GZ CEERS relevant for this paper. Note that you only reach the bar question if you answered `Features or Disc' to the first question and `No - Something Else' to the second question. Please refer to \inprep{Masters et al. (in prep)} for the full decision tree used in GZ CEERS.}
    \label{fig:gz_tree}
\end{figure}

A total of \todo{7,679} colour images were shown to citizen scientists. Every galaxy was classified by at least \todo{40} different volunteers. The entire GZ CEERS project received a total of \todo{311,411} individual classifications. Like its predecessors, GZ CEERS worked with a decision tree structure. The part of the decision tree relevant to this work is shown in Figure \ref{fig:gz_tree}; the full decision tree can be found in \inprep{Masters et al. (in prep)}. Note that volunteers are only asked whether the galaxy has a bar if they previously said that the galaxy has features or a disc and that the galaxy was not viewed edge-on. Volunteers are shown a brief tutorial before they can classify galaxies, which remains accessible at any stage during classification. Importantly for this work, they are also shown examples of strongly and weakly barred galaxies, together with a description of what to look for. We tell them that strong bars are bright, obvious and extend across a large fraction of the galaxy, while weak bars are smaller and fainter.

\subsection{Sample selection}
\label{sec:sample_selection}

Our goal is to measure accurate bar fractions at a range of redshifts. This requires a volume-limited sample of galaxies with robust bar classifications. We describe in this section how this sample is obtained from GZ CEERS.

The initial galaxy detection pipeline was designed to find as many galaxies as possible in the \gls{CEERS} images to include in GZ CEERS. However, the pipeline would occasionally identify one object as two separate galaxies. This was typically more prevalent for irregular galaxies at higher redshifts. This means that there are a significant number of duplicate galaxies in GZ CEERS that have to be removed (see \inprep{Masters et al. (in prep)} for more details). This is done by clustering the targets with an agglomerative clustering algorithm \citep{nielsen_2016}. Each target starts in its own cluster, which are then successively combined together based on the distances between targets. Clusters keep growing in size until a maximum separation of \todo{3} arcsec between targets within each cluster is reached. Note that this method is overly conservative and implies that we likely grouped targets together that are not duplicates, but instead are close companions. One of the authors (DOR) visually inspected each cluster to make sure that was not the case and separated the clusters when necessary. As a final step, another author (TG) removed any leftover duplicates after visually inspecting the cutouts. After this deduplication step, we are left with a sample of \todo{6,640} galaxies. 

We then cross-matched this sample of \todo{6,640} galaxies with the Cosmic Assembly Near-IR Deep Extragalactic Legacy Survey (CANDELS, \citealp{grogin_2011,koekemoer_2011}) in order to use their stellar mass \citep{stefanon_2017} and redshift estimates \citep{kodra_2023}. This is a combination of photometric and spectroscopic redshifts. We use spectroscopic redshifts where available, and supplement with photometric estimates (i.e. we use the \texttt{z\_best} parameter). The normalised median absolute deviation of the differences between the photometric and spectroscopic redshifts in the full CANDELS sample equals 0.0227, and the outlier fraction (defined as the fraction where $ | \Delta z/(1+z) | > 0.15$) equals 0.067 \citep{kodra_2023}. We used a distance threshold of \todo{5} arcsec to match to our sample to ensure accurate cross-matching, which further reduced our sample to \todo{6,135} galaxies.

The next step involves selecting a subset of disc galaxies with reliable bar classifications. In the past (e.g. see \citealt{geron_2021}), this has been done by putting thresholds on the fraction of volunteers that voted that the galaxy has features or a disc ($p_{\textrm{features/disc}}$), the  fraction of volunteers that voted that the galaxy is not viewed edge-on ($p_{\textrm{not edge-on}}$) and the number of volunteers that have been asked the bar question ($N_{\textrm{bar}}$). The classification tree used in GZ CEERS is similar to the one used in GZ DESI. However, as noted by \citet{smethurst_2025}, the distribution of vote fractions in GZ CEERS is different compared to other GZ iterations. This implies that the morphology of these higher redshift ($z \gtrsim 0.5$) galaxies is different compared to the morphology of galaxies in the local Universe. We can therefore not select targets with the same thresholds used in previous GZ iterations. Instead, we have to create new thresholds that are applicable for these high redshift galaxies. Three of the authors of this paper (TG, ILG, DOR) visually inspected the cutouts shown to the volunteers to determine accurate thresholds on $p_{\textrm{features/disc}}$, $p_{\textrm{not edge-on}}$ and $N_{\textrm{bar}}$. Our aim with these thresholds was to be conservative and to select a sample with reliable bar classifications, at the cost of being less complete. We decided on $p_{\textrm{features/disc}}$ $\geq$ \todo{0.3} and $p_{\textrm{not edge-on}}$ $\geq$ \todo{0.54}. These thresholds on $p_{\textrm{features/disc}}$ and $p_{\textrm{not edge-on}}$ effectively ensure that our sample consists of featured or disc galaxies that are not viewed edge-on. However, we still need an additional threshold on $N_{\textrm{bar}}$, as we require sufficient numbers of volunteers to inspect each image in order for the bar classification to be reliable. After visually inspecting the cutouts, the authors decided that $N_{\textrm{bar}}$ $\geq$ \todo{15} is appropriate. Note that applying the $N_{\textrm{bar}}$ thresholds effectively means that the restrictions on $p_{\textrm{features/disc}}$ and $p_{\textrm{not edge-on}}$ become even more restrictive. Note that using slightly different threshold values does not significantly change the results of this work. Applying these three thresholds reduced our sample size to \todo{668} galaxies.

The \todo{668} disc galaxies in this sample all have reliable classifications for the bar question (though not all of them have bars). We will call this sample the `parent' sample. However, because we are interested in studying bar fractions across redshift, we need to make a volume-limited sample in order to avoid selection biases. \gls{CEERS} has a 5$\sigma$ limiting magnitude of 28.3 - 29.2, depending on which band is used \citep{kauffmann_2020, bagley_2023, finkelstein_2023}. We summed the fluxes in all bands and converted this to a combined magnitude in order to not be sensitive to any particular filter. GZ CEERS only showed galaxies to volunteers if the combined apparent magnitude was $m < 24$ \inprep{(Masters et al. in prep)}, which is much brighter than the limiting magnitudes cited above. Additionally, other studies have previously used a magnitude threshold of 27 to select galaxies with reliable morphological classifications in \gls{CEERS} \citep{kartaltepe_2023, huertas_company_2024}. To make the sample volume-limited, we constrained it to galaxies with redshifts $0.5 < z < 4$ and absolute magnitude \todo{$M < -23.77$}. This lowered our sample size down to \todo{398} galaxies. The multiple thresholds used to create this sample, together with sample size at every step, are summarised in Table \ref{tab:sample_selection}. We will use the volume-limited sample throughout this work, unless it is explicitly mentioned that we use the parent sample instead.

\begin{deluxetable}{c c}
  \tablecaption{Summary of the sequential sample selection used to create the volume-limited sample used in this work. We also show the sample size next to every step.}
  \label{tab:sample_selection}
  \tablehead{
  \colhead{Threshold} & \colhead{Sample size}}
  \decimalcolnumbers
  \startdata
  GZ CEERS & 7,679 \\
  Deduplication & 6,640 \\
  Match to CANDELS & 6,135 \\
  $p_{\textrm{features/disc}}$ $>$ 0.3 & 1,952 \\
  $p_{\textrm{not edge-on}}$ $>$ 0.54 & 1,277 \\
  $N_{\textrm{bar}}$ $\geq$ 15 & 668\\
  Volume-limit & 398\\
  \enddata
\end{deluxetable}

\begin{figure}
	\includegraphics[width=\columnwidth]{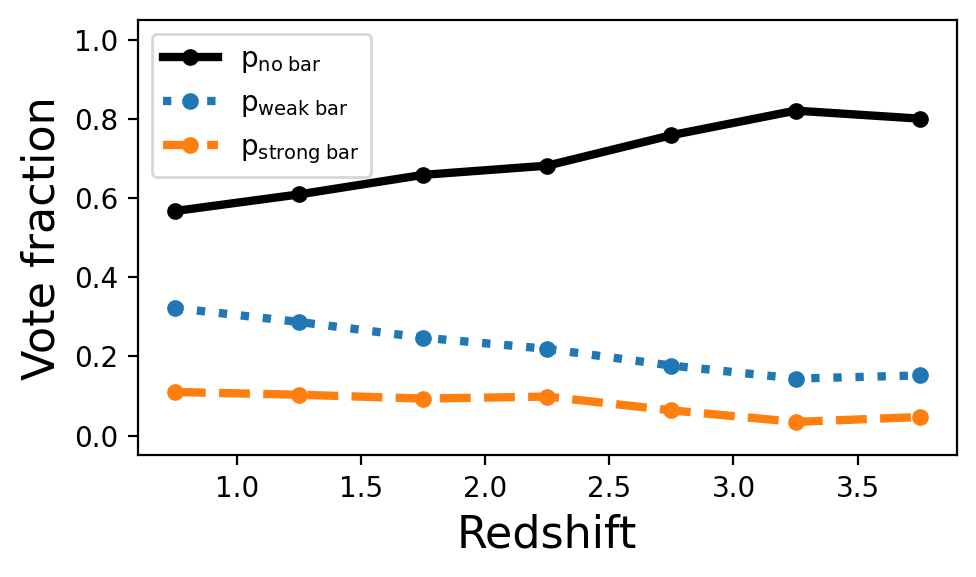}
    \caption{The average fraction of people that voted that the galaxy is unbarred ($p_{\textrm{no bar}}$, black), weakly barred ($p_{\textrm{weak bar}}$, blue) and strongly barred ($p_{\textrm{strong bar}}$, orange) plotted against redshift. Note that this does not show the actual bar fraction, but the GZ vote fraction in every redshift bin. The method used to convert from GZ vote fractions to bar classifications is detailed in Section \ref{sec:sample_selection}. The weak bar vote fraction clearly decreases with redshift, while the unbarred vote fraction increases. The strong bar vote fractions remains roughly constant.}
    \label{fig:vote_fractions}
\end{figure}

\begin{figure*}
	\includegraphics[width=\textwidth]{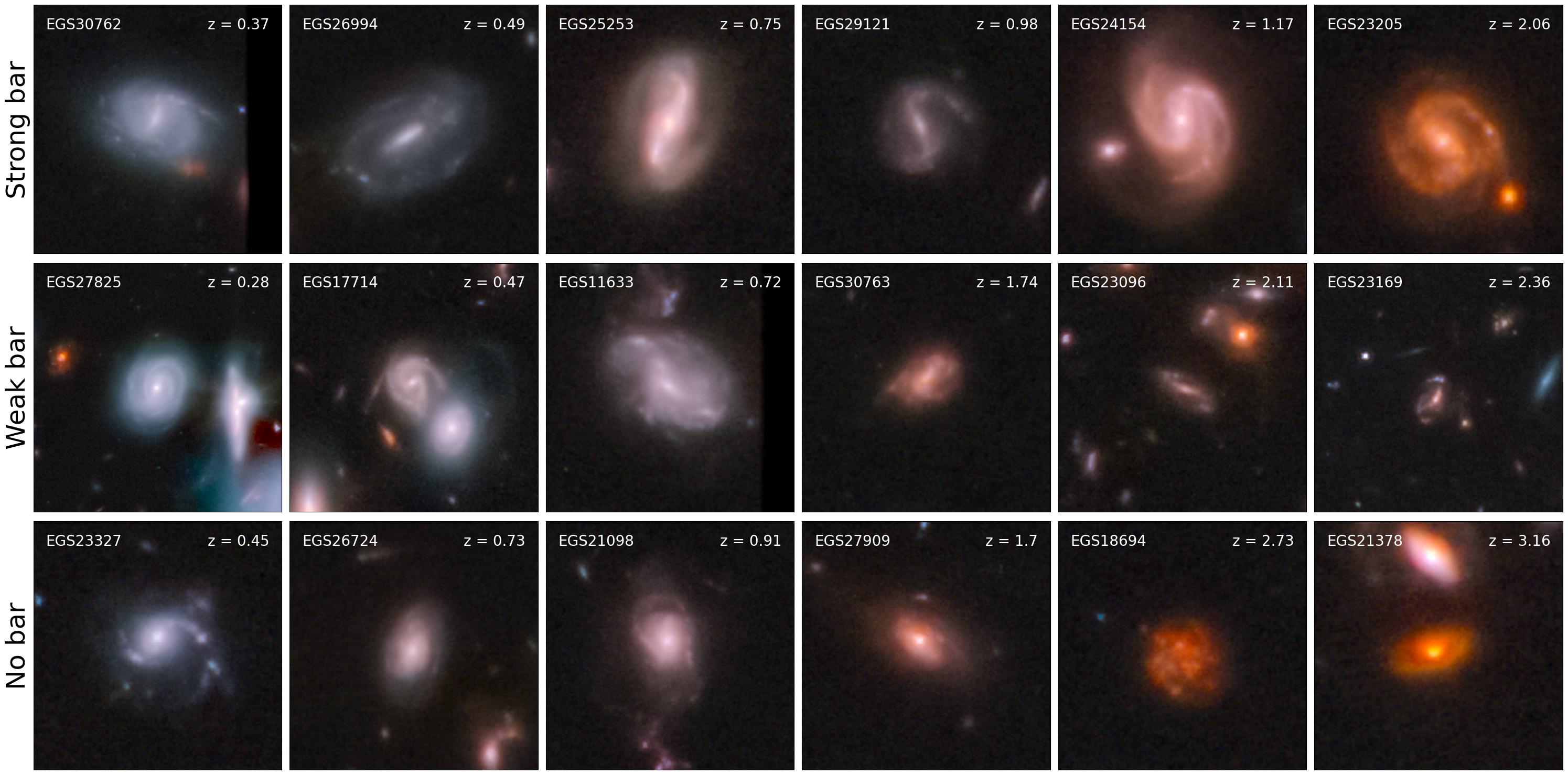}
    \caption{A collection of coloured \gls{CEERS} cutouts of strongly barred (top row), weakly barred (middle row) and unbarred (bottom row) galaxies found in the parent sample. The colouring matches how the images were presented to the volunteers. A full list of the barred galaxies can be found in Appendix \ref{app:all_images}. The CANDELS ID of each galaxy is shown in the top left corner of each image, while the redshift is shown in the top right corner.}
    \label{fig:overview}
\end{figure*}

The bar question has three possible answers: strong bar, weak bar and no bar. The three associated vote fractions are: $p_{\textrm{strong bar}}$, $p_{\textrm{weak bar}}$ and $p_{\textrm{no bar}}$. The evolution of these vote fractions in the volume-limited sample over redshift is shown in Figure \ref{fig:vote_fractions}. It is clear that the weak bar vote fraction decreases with redshift, while the unbarred vote fraction increases. The strong bar vote fraction remains roughly constant over redshift. Note that Figure \ref{fig:vote_fractions} does not show the bar fractions, but rather the vote fractions. We can assign a bar type (strong, weak or no bar) to every galaxy using the vote fractions based on the classification scheme detailed below. A galaxy is classified as unbarred if more than half of all the classifications voted that the galaxy did not have a strong or weak bar, i.e. $p_{\textrm{strong bar + weak bar}} < 0.5$. If this was not the case, then the galaxy had a strong bar if $p_{\textrm{strong bar}} \geq p_{\textrm{weak bar}}$ and a weak bar if $p_{\textrm{strong bar}} < p_{\textrm{weak bar}}$. Using this scheme, we find a total of \todo{161} bars in the parent sample, of which \todo{35} are strong bars and \todo{126} are weak bars. Similarly, we find a total of \todo{87} bars in the volume-limited sample, of which \todo{17} are strong bars and \todo{70} are weak bars. This bar strength classification scheme has been successfully used before in other work (e.g. see \citealt{geron_2021, geron_2023, garland_2024}) using data from GZ DECaLS and GZ DESI. After inspecting the resultant cutouts, it was concluded that the same bar strength classification scheme can be used at these higher redshifts as well. Figure \ref{fig:overview} shows a few examples of strongly barred, weakly barred and unbarred galaxies found in the parent sample of GZ CEERS. Cutouts of all strongly barred and weakly barred galaxies in the parent sample can be found in Appendix \ref{app:all_images}.

\begin{figure}
	\includegraphics[width=\columnwidth]{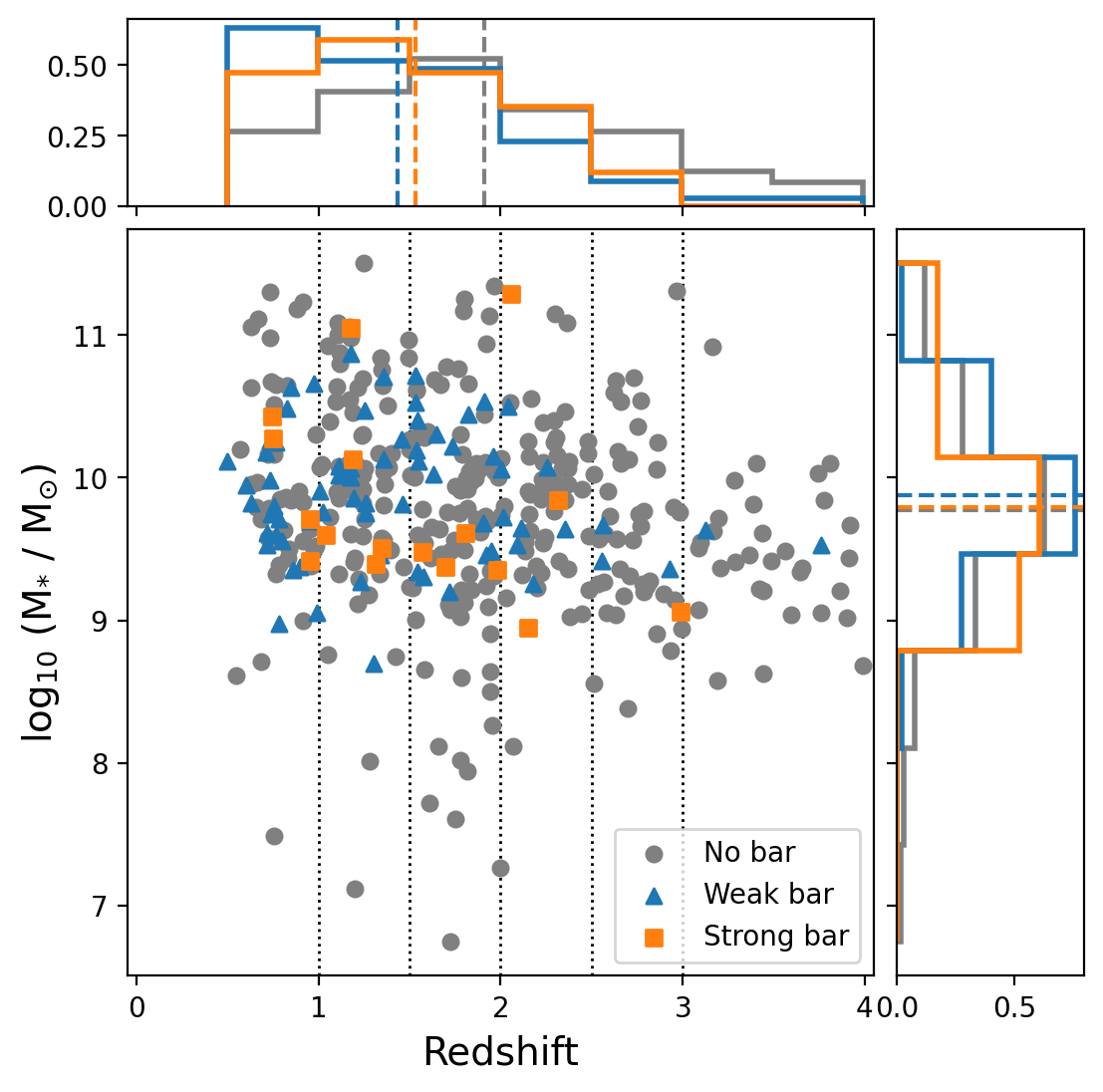}
    \caption{Redshift against stellar mass (both obtained from CANDELS, \citealt{stefanon_2017, kodra_2023}) for the strongly barred (orange), weakly barred (blue) and unbarred (grey) galaxies in our volume-limited sample. The vertical dotted lines in the central panel represent the edges of the redshift bins used in this work. The dashed lines in the histograms denote the median values for the redshift and stellar mass distributions for the different bar types.}
    \label{fig:mass_z}
\end{figure}

Although we do make our sample volume-limited to enable comparisons of bar fractions across redshift, we want to double check that our sample is not biased towards lower masses in lower redshift bins. This is done in Figure \ref{fig:mass_z}, which shows redshift against stellar mass (both obtained from CANDELS, \citealt{stefanon_2017, kodra_2023}) for the strongly barred (orange), weakly barred (blue) and unbarred (grey) galaxies in our volume-limited sample. We find that, based on an Anderson-Darling test, the stellar mass distribution between strongly barred, weakly barred and unbarred galaxies in the volume-limited sample is not significantly different ($<2\sigma$). We also note that our volume-limited sample is not dominated by lower stellar masses in the lower redshift bins. However, there are a few galaxies with lower stellar masses ($<10^{9} M_{\odot}$) in the bins up to $z = 3$, but not as many in the highest redshift bin. As a test, we removed these galaxies, and found that it does not significantly change any of our results. This confirms that our volume-limited sample behaves as intended.

\subsection{Calculating bar fractions}
\label{sec:bar_fraction}

In this work, whenever we use the term `bar fraction', we mean `observed bar fraction', and not `true bar fraction'. The bar fraction, $f_{\rm bar}$, can be calculated as:

\begin{equation}
    f_{\rm bar} = \frac{N_{\rm bar}}{N_{\rm disc}}\;,
    \label{eq:barfrac}
\end{equation}
where $N_{\rm bar}$ is the number of barred galaxies and $N_{\rm disc}$ is the number of discs. However, there are multiple corrections that we can make to the bar fraction to remove sources of bias. In Section \ref{sec:corrections_fd}, we discuss how to more accurately estimate the denominator of Equation \ref{eq:barfrac} to include potentially missed featureless discs. We discuss redshift corrections in Section \ref{sec:corrections_z}. We opt for the most minimal correction factors when calculating the final corrected bar fraction, which means that these corrected values should be interpreted as lower limits and that the ``true'' bar fraction likely lies above the corrected bar fractions quoted here. We share the raw bar fractions, as well as every combination of corrected bar fractions in Table \ref{tab:barfracs}. A machine-readable version of the bar fractions determined in this work can be found \href{https://github.com/tobiasgeron/gz_ceers_bar_fractions}{here}\footnote{\url{https://github.com/tobiasgeron/gz_ceers_bar_fractions}}. Finally, we discuss how we obtain error bars on the bar fractions in Section \ref{sec:error_bars}.

\subsubsection{Correcting for featureless discs}
\label{sec:corrections_fd}

A bar fraction can be defined as the number of barred galaxies ($N_{\rm bar}$) over the number of disc galaxies ($N_{\rm disc}$). The number of barred galaxies can be estimated using GZ CEERS, as described above. We can estimate the number of disc galaxies using GZ CEERS as well, as the first question volunteers answer is whether they see features or a disc (see Figure \ref{fig:gz_tree}). However, previous Galaxy Zoo iterations have shown that volunteers are more sensitive to picking up features rather than discs when answering this question. Featureless discs, lenticular galaxies and S0 galaxies are typically classified as ``smooth'' \citep{willett_2013, simmons_2017,walmsley_2023}. This implies that using GZ CEERS to select discs will bias you towards featured discs, and that Equation \ref{eq:barfrac} becomes:

\begin{equation}
    f_{\rm bar, raw} = \frac{N_{\rm bar}}{N_{\rm featured}}\;,
    \label{eq:barfrac_gz}
\end{equation}
where $f_{\rm bar, raw}$ is the raw bar fraction. As featureless discs, lenticulars and S0 galaxies are relatively rare, this was not a major issue and previous papers could assume that $N_{\rm disc} \approx N_{\rm featured}$ (e.g. see \citealt{masters_2011,melvin_2014,geron_2021}). However, recent work by \citet{smethurst_2025} has shown that this is no longer the case for GZ CEERS. Using the disc classifications from \citet{ferreira_2023}, they have found a significant number of featureless discs at these higher redshifts. As these discs are classified as ``smooth'' by the volunteers, they are not included in the denominator of the bar fractions calculated above. Thus, a correction to the denominator of Equation \ref{eq:barfrac_gz} is needed.

We can do this by considering all galaxies that did not pass the featured threshold ($p_{\textrm{features/disc}}$) discussed in Section \ref{sec:sample_selection} and making that sample also volume-limited. These \todo{829} galaxies are all classified as having no features (and no disc) in GZ CEERS, which we will call the `featureless' sample. We can double check whether these featureless galaxies actually have no disc using the disc classifications of \citet{ferreira_2023}. Here, six experts visually classified the galaxies found in \gls{CEERS} into five different categories: unclassifiable, point source, disc, spheroid and peculiar. If more than half (i.e. 4 out of 6) experts said a source was a disc, then it was classified as a disc galaxy. However, \citet{ferreira_2023} only classified galaxies with $z > 1.5$. There are \todo{678} galaxies in the featureless sample that are $z > 1.5$, of which \todo{534} are classified by \citet{ferreira_2023}. We find that \todo{206} of these \todo{534} galaxies (or \todo{39\%}) have discs according to \citet{ferreira_2023}, despite having been classified as featureless (and disc-less) in GZ CEERS. These are the featureless discs that we need to account for in our bar fractions. For a more detailed study on these featureless discs, please refer to \citet{smethurst_2025}. Thus, we can adjust the denominator of Equation \ref{eq:barfrac_gz} and include these featureless discs and calculate a bar fraction that accounts for featureless discs ($f_{\rm bar, FD}$) with: 

\begin{equation}
    f_{\rm bar, FD} = \frac{N_{\rm bar}}{N_{\rm featured} + N_{\rm featureless\;discs}}\;,
    \label{eq:barfrac_f23}
\end{equation}
where $N_{\rm featureless\;discs}$ is the number of featureless discs identified in the featureless sample. Since $N_{\rm featureless\;discs} \geq 0$, applying this threshold will lower the overall bar fraction. We compute this new bar fraction for every redshift bin.

There are a few caveats to this approach that we need to address. Firstly, \todo{144} galaxies in our featureless sample with $z > 1.5$ do not have disc classifications in \citet{ferreira_2023}. However, we can mitigate this by assuming that the fraction of featureless discs for these galaxies is similar to the ones that do have disc classifications. Secondly, as mentioned above, \citet{ferreira_2023} only consider galaxies with $z > 1.5$. To deal with this problem, we assume that the galaxy population at $z < 1.5$ has a similar ratio of featureless discs as at $z = 1.5$, and use a similar correction factor for these lower redshift bins. This is possibly an overcorrection, which implies that we would underestimate the final corrected bar fraction at these lower redshift bins. This also means that it is more appropriate to interpret these corrected bar fractions as lower limits.

\subsubsection{Redshift correction}
\label{sec:corrections_z}

A major concern for any study that looks at bar fractions is whether all bars are actually detected. This is especially an issue at higher redshifts, where resolution and cosmological dimming effects can obscure bars. We show in Appendix \ref{app:resolution} that resolution effects alone cannot be responsible for hiding a significant number of bars in JWST, as the \gls{FWHM} of the bands, especially the bluer bands, is sufficiently small.

Another effect that can artificially lower the bar fraction is cosmological surface brightness dimming, which scales as $(1+z)^{-4}$. However, the intrinsic galaxy surface brightness has been found to brighten with redshift, which is typically explained by a combination of luminosity evolution and size evolution of the galaxies themselves \citep{barden_2005, sobral_2013, yu_2023}. This suggests that cosmological surface brightness dimming is not a limiting factor for detecting bars with \gls{HST} up to $\sim$1 \citep{sheth_2008}, although \citet{whitney_2020} argues that this is an observational bias and we only find this effect because we miss low surface brightness galaxies at higher redshifts. Either way, the detection fraction of bars in \gls{CEERS} was investigated in greater detail by \citet{yu_2023} and \citet{liang_2024}. \citet{yu_2023} defined a sample of 1,816 nearby galaxies observed in the \gls{DESI} survey and artificially redshifted them up to redshift of 3 to generate simulated \gls{CEERS} images. The procedure for simulating these images included accounting for both observational effects (e.g. lower resolution, less signal and more noise) and galaxy evolution effects (e.g. physical bar and disc size evolution and luminosity evolution). A subset of these galaxies (N = 448) were used in \citet{liang_2024} to simulate the detection fraction of bars across a redshift range of 0 $<$ $z$ $<$ 3. The true bar fraction in the artificially redshifted sample was equal to 68\%, though they found that the detected bar fraction was often much lower than that, depending on the redshift and band used. This implies that the measured bar fraction in the simulations of \citet{liang_2024} is decreasing with redshift, not because of changes in the true underlying bar fraction, but because of other observational biases which should be corrected for. 

\begin{figure*}
	\includegraphics[width=\textwidth]{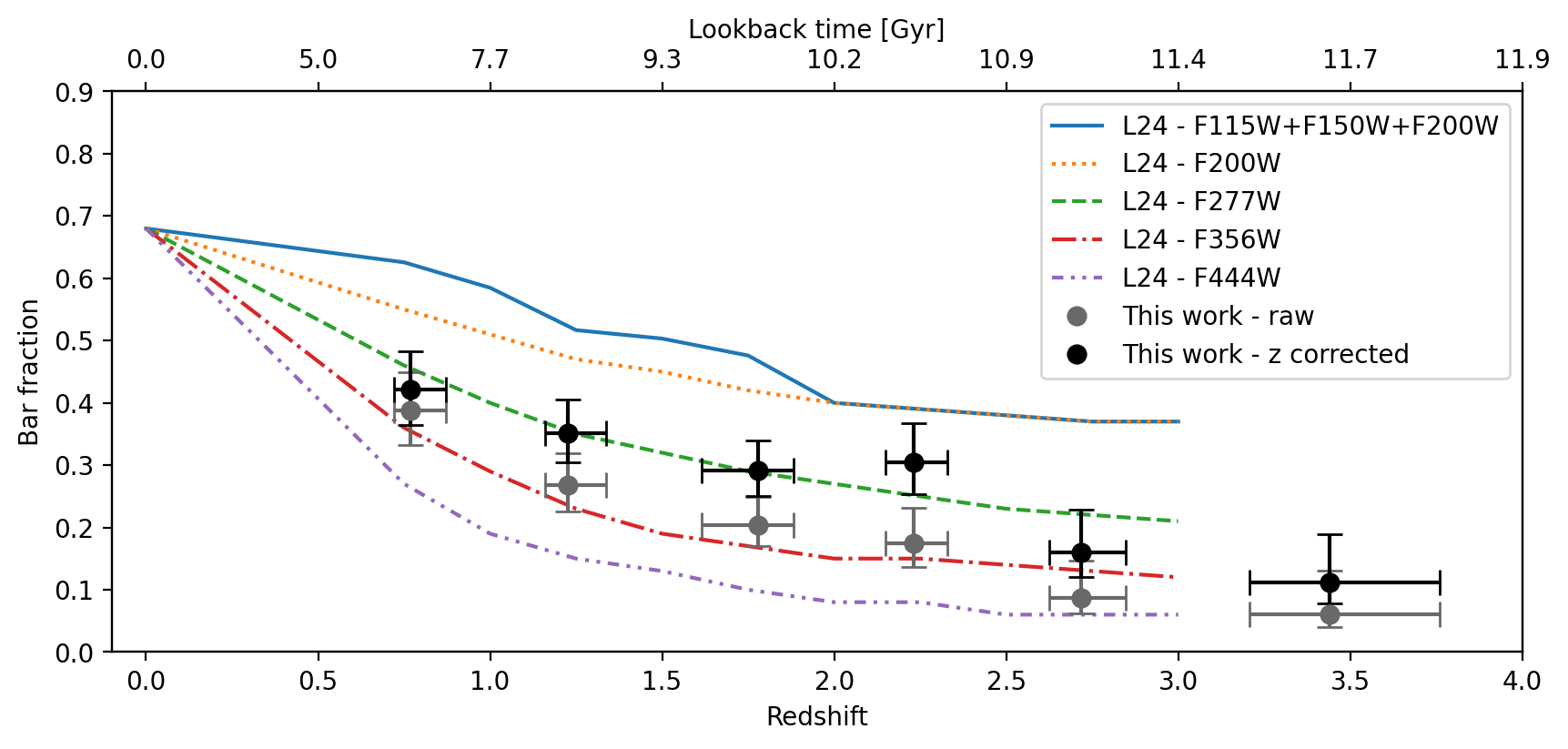}
    \caption{The bar fractions fraction measured in the simulations of \citet{liang_2024} for different bands over redshift (see their Table 3 and Figure 15). They kept the true bar fraction constant at 68\%, but found that the measured bar fraction decreased with redshift due to observational biases. The rate of decrease depends on which band is considered. In this work, we correct for redshift by using a combination of the detection fractions of the F115W, F150W and F200W bands, shown here in blue. We also show our uncorrected bar fractions (grey dots), as well as the bar fractions corrected for redshift (black dots). Note that these corrected bar fractions are only corrected for redshift, the featureless disc correction is not shown here.}
    \label{fig:z_correction}
\end{figure*}

The evolution of the simulated detected bar fraction found in \citet{liang_2024} for the F200W, F277W, F356W and F444W bands is shown in Figure \ref{fig:z_correction}, together with the uncorrected raw bar fractions found in this work. Note that the decrease in our observed bar fraction is similar to the decrease found in the simulated bar fractions in the F356W band of \citet{liang_2024}. Unfortunately, \citet{liang_2024} only measured the detection fraction for single-band images, while we use colour images to find bars in this work, which makes a direct comparison difficult. However, this implies that, if the decrease with redshift in our multi-band coloured images is similar to the decrease in the simulated measured bar fractions of the F356W band, the true bar fraction found here is consistent with being constant. It is unlikely that this is the case, but interesting to note nonetheless. 

As mentioned above, it is hard to directly compare these single-band simulations with our coloured images. Nevertheless, we still want to apply a redshift correction to mitigate redshift effects. Thus, we opted for the most minimal redshift correction. This is a correction obtained from a combination of the detection fractions of the F115W, F150W, and F200W bands in the simulations of \citet{liang_2024}. As they note, while F115W had the lowest \gls{FWHM} and therefore can detect the smallest structures, it will also miss bars at higher redshifts due to its shifting to shorter rest-frame wavelengths. Thus, \citet{liang_2024} simulate the effectiveness of detecting bars using a combination of these three bands. F115W is used for $z$ $\leq$ 1, F150W for 1 $<$ $z$ $<$ 2, and F200W for $z$ $\geq$ 2. These fractions can be found in Figure 15 of \citet{liang_2024} and their inverses are used as correction factors in this work going forward. This combined detection fraction and the bar fraction corrected for redshift effects ($f_{\rm bar, z}$) using this correction factor are also shown in Figure \ref{fig:z_correction}. We can apply this correction factor to Equation \ref{eq:barfrac_f23} to calculate a bar fraction that accounts for both featureless discs and redshift effects ($f_{\rm bar, FD+z}$):

\begin{equation}
    f_{\rm bar, FD+z} = \frac{N_{\rm bar}}{N_{\rm featured} + N_{\rm featureless\;discs}} \cdot k\;,
    \label{eq:barfrac_corr}
\end{equation}
where $k$ is the redshift correction. This increases the bar fraction across all redshifts. We highlight that this is the most minimal realistic redshift correction we can apply. It is likely that the correct correction factor is higher, but it is hard to estimate without performing more sophisticated simulations with coloured images, which is outside the scope of this work.  This also means that our corrected bar fractions are best interpreted as lower limits.

Note that the simulations from \citet{liang_2024} only go up to $z$ $=$ 3. We use the correction factor of the $z$ $=$ 3 simulations for observations with $z$ $>$ 3 in order to be conservative and not extrapolate the correction factor ourselves.

\subsubsection{Estimating uncertainty}
\label{sec:error_bars}

The statistical uncertainty on the bar fraction is estimated using the beta distribution quantile technique, as described in \citet{cameron_2011}. They show that this method of calculating the confidence intervals is more robust than other popular techniques, such as the `normal approximation' or the method developed by \citealt{clopper_1934}. The former underestimates the confidence intervals when $p$ approaches 0 or 1, especially for smaller sample sizes, while the latter overestimates the confidence intervals. Whether a random galaxy is barred is a Bernoulli trial, which implies that the total bar fraction of an entire sample follows the binomial distribution. This, in turn, defines a beta distribution $B(a,b)$, with parameters $a = N_{\rm bar} + 1$ and $b = N_{\rm disc} - N_{\rm bar} + 1$. Using the beta distribution quantile technique, we can then find the upper ($p_{u}$) and lower bounds ($p_{l}$) for any confidence level $c=1-\alpha$ with:

\begin{equation}
    \int^{p_{l}}_{0} B(a,b) dp = \alpha/2 \;\; \textrm{and} \;\; \int^{1}_{p_{u}} B(a,b) dp = \alpha/2 \;.
\end{equation}
Refer to \citet{cameron_2011} for more details on this topic. This technique to quantify the uncertainty on bar fractions was also previously used in \citet{simmons_2014} and \citet{leconte_2024}. We use $c = 0.68$ in this work.

\section{Results}
\label{sec:results}

We apply two different corrections to the raw bar fractions in this work: a correction to include featureless discs (see Section \ref{sec:corrections_fd}) and a correction to account for redshift effects (see Section \ref{sec:corrections_z}). The former lowers the bar fraction, while the latter increases the bar fraction. However, across all redshift bins, the featureless disc correction is stronger than the redshift correction. This means that the bar fraction corrected for both effects is lower than the raw bar fraction. To be explicit on how each individual correction affects the bar fraction across redshift, we share the bar fractions of all possible combinations in Table \ref{tab:barfracs}. In the remainder of this work, when we mention corrected bar fractions, we refer to bar fraction adjusted for both redshift and featureless discs, unless stated otherwise.

\begin{figure*}
    \centering
	\includegraphics[angle = 90, width=0.80\textwidth]{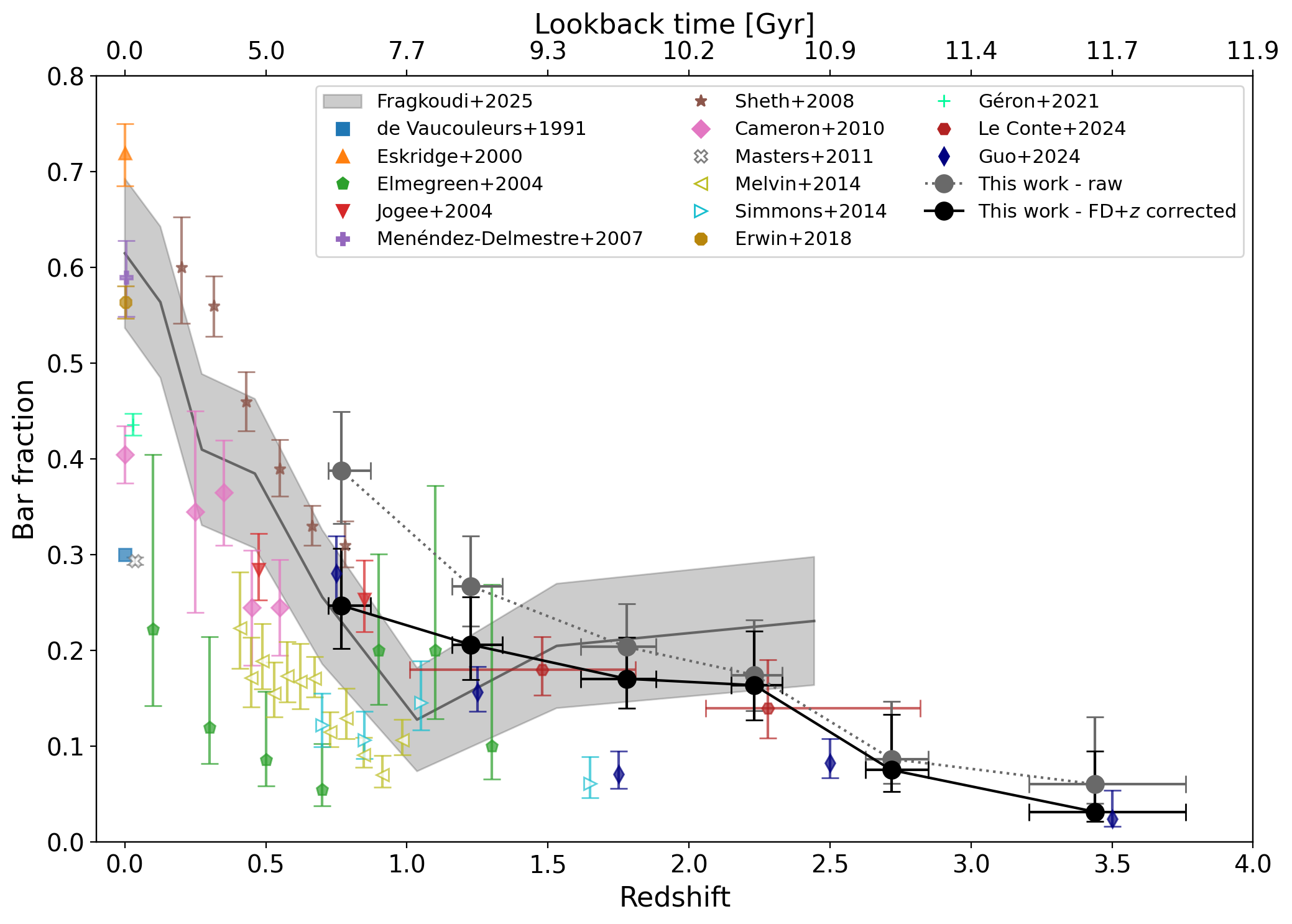}
    \caption{The bar fraction found in this work plotted over redshift (grey and black dots). We show both the raw, uncorrected bar fractions (dotted line) and corrected bar fractions (solid line). We see a general downward trend of bar fraction with redshift. Data from other studies are included in this figure to facilitate comparison. Some of these studies count both weak and strong bars (filled markers), while others focussed on mostly strong bars (open markers). We include observations from \citet{devaucouleurs_1991,eskridge_2000, elmegreen_2004,jogee_2004,menendez_delmestre_2007, sheth_2008, cameron_2010, masters_2011, melvin_2014, simmons_2014, erwin_2018, geron_2021, leconte_2024,guo_2024}. We also show the simulations of \citet{fragkoudi_2025} in the shaded regions. Considering the errors, we are in broad agreement with previous observations at higher redshifts and with the simulation of \citet{fragkoudi_2025}.}
    \label{fig:main}
\end{figure*}

The raw and corrected total bar fractions are plotted against redshift in Figure \ref{fig:main}. There is an overall decreasing trend with redshift; the raw bar fraction starts at \todo{$39^{+6}_{-6}$}\% at $0.5 < z < 1.0$ and goes down to \todo{$6^{+7}_{-2}$}\% at $3.0 < z < 4.0$. Similarly, the corrected bar fraction starts at \todo{$25^{+6}_{-4}$}\% and goes down to \todo{$3^{+6}_{-1}$}\%. This is a $>$$3\sigma$ significant decrease between the lowest and highest redshift bin for both the raw and corrected bar fractions. We have added bar fractions from other observations \citep{devaucouleurs_1991,eskridge_2000, elmegreen_2004,jogee_2004,menendez_delmestre_2007, sheth_2008, cameron_2010, masters_2011, melvin_2014, simmons_2014, erwin_2018, geron_2021, leconte_2024,guo_2024} and the simulation of \citet{fragkoudi_2025} in this plot to facilitate comparison with other work. Our corrected bar fractions agree well with the simulations of \citet{fragkoudi_2025}. At lower redshifts ($0.5 < z < 1.0$), our corrected bar fractions also agree with \citet{elmegreen_2004,jogee_2004, sheth_2008}. Our corrected bar fractions are also consistent with \citet{leconte_2024}, a previous study looking at bar fractions in CEERS across all redshifts. Our corrected bar fractions are also consistent with bar fractions of \citet{guo_2024} in the low redshift and high redshift bins, though we find higher values for the bar fraction in the middle redshift bin ($ 1.5 < z < 2.0$). This is most likely because we apply a redshift correction, while they do not. A more detailed comparison to other work can be found in Section \ref{sec:disc_barfrac}.

\begin{deluxetable*}{c c c c c c c c c}
    \tablecaption{A summary of the bar fractions and sample size in each redshift bin for the volume-limited sample. We show the uncorrected raw bar fraction ($f_{\rm bar, raw}$), the bar fraction that includes featureless discs in the denominator ($f_{\rm bar, FD}$), the bar fraction that applies a redshift correction ($f_{\rm bar, z}$) and a bar fraction that does both ($f_{\rm bar, FD+z}$).This table can be found in machine-readable format \href{https://github.com/tobiasgeron/gz_ceers_bar_fractions}{here}.}
    \label{tab:barfracs}
    \tablehead{
    \colhead{Redshift} & \colhead{$f_{\rm bar, raw}$} & \colhead{$f_{\rm bar, FD}$} & \colhead{$f_{\rm bar, z}$} & \colhead{$f_{\rm bar, FD+z}$} & \colhead{Sample size} & \colhead{Number of bars} & \colhead{Number of strong bars} & \colhead{Number of weak bars} 
    }
    \decimalcolnumbers
    \startdata
        0.5 $< z <$ 1.0 & 0.39$^{ +0.06 }_{ -0.06 }$ & 0.23$^{ +0.06 }_{ -0.04 }$ & 0.42$^{ +0.06 }_{ -0.06 }$ & 0.25$^{ +0.06 }_{ -0.04 }$ & 67 & 26 & 4 & 22\\ 
        1.0 $< z <$ 1.5 & 0.27$^{ +0.05 }_{ -0.04 }$ & 0.16$^{ +0.05 }_{ -0.03 }$ & 0.35$^{ +0.05 }_{ -0.05 }$ & 0.21$^{ +0.05 }_{ -0.04 }$ & 86 & 23 & 5 & 18\\ 
        1.5 $< z <$ 2.0 & 0.20$^{ +0.04 }_{ -0.03 }$ & 0.12$^{ +0.04 }_{ -0.02 }$ & 0.29$^{ +0.05 }_{ -0.04 }$ & 0.17$^{ +0.04 }_{ -0.03 }$ & 103 & 21 & 4 & 17\\ 
        2.0 $< z <$ 2.5 & 0.17$^{ +0.06 }_{ -0.04 }$ & 0.09$^{ +0.05 }_{ -0.02 }$ & 0.30$^{ +0.06 }_{ -0.05 }$ & 0.16$^{ +0.06 }_{ -0.04 }$ & 63 & 11 & 3 & 8\\ 
        2.5 $< z <$ 3.0 & 0.09$^{ +0.06 }_{ -0.03 }$ & 0.04$^{ +0.05 }_{ -0.01 }$ & 0.16$^{ +0.07 }_{ -0.04 }$ & 0.08$^{ +0.06 }_{ -0.02 }$ & 46 & 4 & 1 & 3\\ 
        3.0 $< z <$ 4.0 & 0.06$^{ +0.07 }_{ -0.02 }$ & 0.02$^{ +0.06 }_{ -0.00 }$ & 0.11$^{ +0.08 }_{ -0.03 }$ & 0.03$^{ +0.06 }_{ -0.01 }$ & 33 & 2 & 0 & 2\\ 
    \enddata
\end{deluxetable*}

\begin{figure*}
	\includegraphics[width=\textwidth]{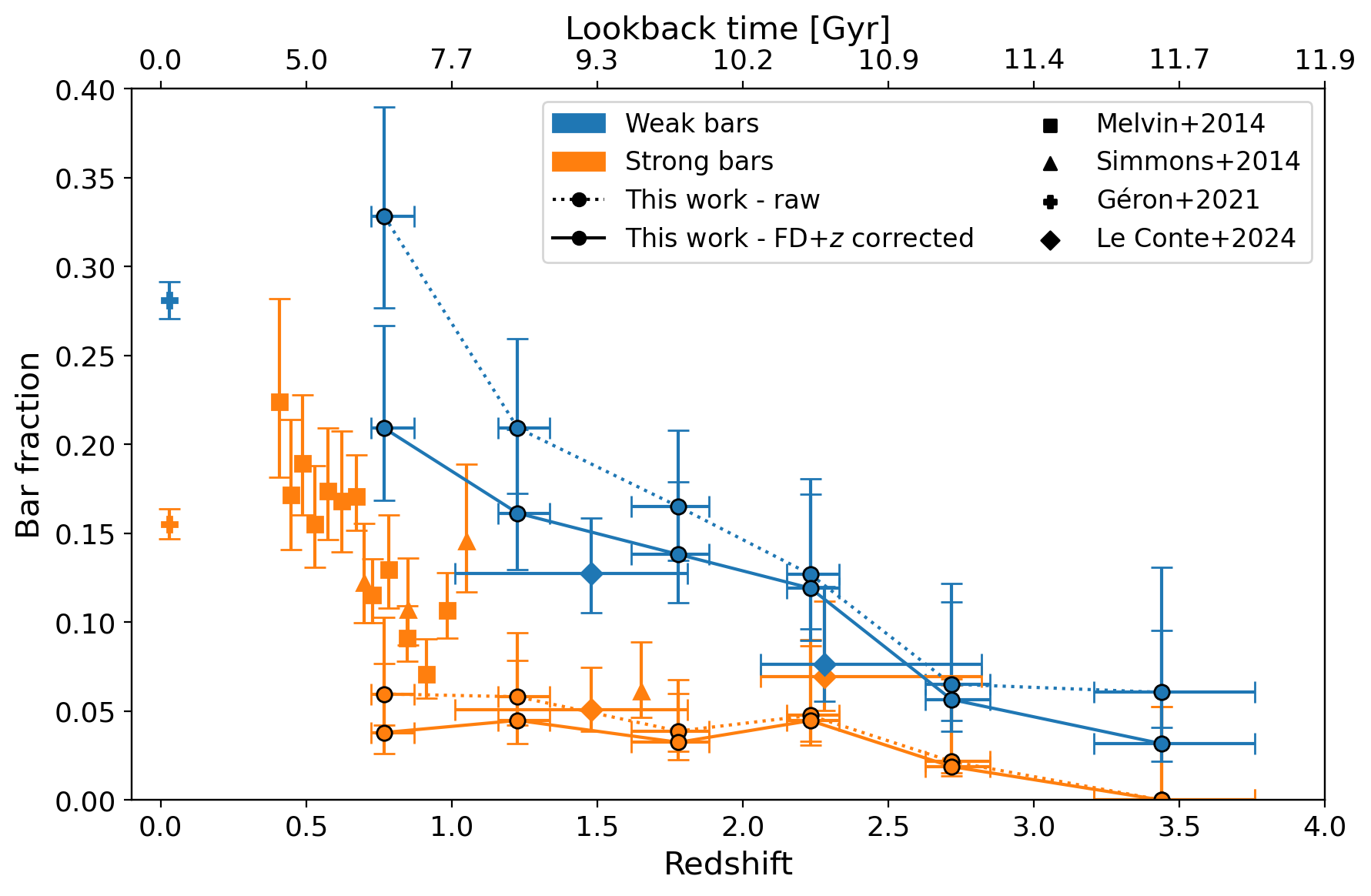}
    \caption{The evolution of the strong and weak bar fraction over redshift. The orange lines show the strong bar fraction and the blue lines show the weak bar fraction. The dotted lines represent the raw, uncorrected bar fractions (see Equation \ref{eq:barfrac}), while the solid lines represent the corrected bar fractions (see Equation \ref{eq:barfrac_corr}). We also show strong and weak bar fractions of other observations that explicitly distinguish between weak and strong bars \citep{melvin_2014,simmons_2014, geron_2021, leconte_2024}. The fractions obtained in this work are additionally outlined in black. The strong bar fraction seems to evolve less over redshift than the weak bar fraction.}
    \label{fig:bar_strength}
\end{figure*}

We look at the raw and corrected weak and strong bar fraction over redshift in Figure \ref{fig:bar_strength}. We also plot the bar fractions of other observations that explicitly distinguish between strong and weak bar \citep{melvin_2014,simmons_2014, geron_2021, leconte_2024}. The weak bar fraction seems to be decreasing significantly with redshift in this work, both in the raw and corrected fractions. The corrected weak bar fraction starts at \todo{$21^{+6}_{-4}$}\% at $0.5 < z < 1.0$ and goes down to \todo{$3^{+6}_{-1}$}\% at $3.0 < z < 4$. Interestingly, the corrected strong bar fraction remains roughly constant around \todo{4\%} between $0.5 < z < 2.5$. Both our corrected weak and strong bar fractions are consistent within the error bars of the weak and strong bar fractions reported in \citet{leconte_2024}. However, the strong bar fraction at the lowest bin ($0.5 < z < 1$) is considerably lower here than what was found in $z \approx 0$ studies that look at strong bar fractions \citep{geron_2021}. This implies that there is a large evolution in the strong bar fraction between $z = 0$ and our lowest redshift bin, which is supported by the observations of \citet{melvin_2014}. Alternatively, the vote fractions might also be affected by unknown observational effects. The implications of these strong and weak bar trends are discussed in greater detail in Section \ref{sec:disc_bar_strength}.

\section{Discussion}
\label{sec:discussion}

\subsection{Bar fraction at high redshifts}
\label{sec:disc_barfrac}

We show the raw and corrected bar fractions found in this work between $0.5 < z < 4.0$ in Figure \ref{fig:main}. As explained in Section \ref{sec:bar_fraction}, we applied a redshift correction and accounted for featureless discs when calculating the corrected bar fractions. However, we used the most minimal corrections, which implies that the corrected bar fractions should be interpreted as lower limits and that the true bar fraction lies higher. We also plot bar fractions from other observations (e.g. \citealp{jogee_2004,sheth_2008,erwin_2018}) and the simulations of \citet{fragkoudi_2025} in Figure \ref{fig:main}. We find an overall decrease in both the raw and corrected bar fractions. This is in agreement with the simulations of \citet{fragkoudi_2025}. Our bar fractions seem to disagree with the simulations of \citet{rosas_guevara_2022}, who predict a mostly constant bar fraction of 40\% between 0.5 $\geq$ $z$ $\geq$ 3.0. However, \citet{rosas_guevara_2022} mention that their simulations are consistent with other observations, such as \citet{sheth_2008} and \citet{melvin_2014}, when they account for the limited spatial resolution of these observations. While we did apply a redshift correction that accounts for resolution effects, it is likely that a similar issue is the reason for the discrepancy with this work, especially since we used the most minimal correction available (see Section \ref{sec:corrections_z}). We expect our results to become more consistent with the simulations of \citet{rosas_guevara_2022} if a stronger redshift correction is used. A key difference between the results of the simulations \citet{rosas_guevara_2022} and \citet{fragkoudi_2025} is that the former predicts a constant bar fraction with a lot of smaller bars ($<$2 kpc) that observations will miss at higher redshifts, while the latter predicts a decreasing bar fraction. Even with the superior resolution and depth of JWST, it is still hard to distinguish between these two possibilities.

At lower redshifts ($0.5 < z < 1.0$), our corrected bar fractions agree well with other observations, such as \citet{elmegreen_2004,jogee_2004, sheth_2008}. We do find higher bar fractions than \citet{melvin_2014} and \citet{simmons_2014}, who used \gls{HST} between $0.5 < z < 2.0$, although that is likely because these studies only focussed on stronger bars. The two main other observational studies we want to compare with are \citet{leconte_2024} and \citet{guo_2024}, as they both also examined bar fractions in \gls{CEERS}. \citet{guo_2024} find a total of 111 barred galaxies with their visual classifications in either the F200W or F444W bands. \citet{leconte_2024} find 39 bars in JWST using the F444W band. We find \todo{161} bars in our parent sample, the largest number of barred galaxies found in the high redshift Universe to date, of which \todo{87} remain in the volume-limited sample. However, the two other studies identify bars in a slightly different way compared to us. Firstly, \citet{leconte_2024} identify bars by visually inspecting galaxies in the F444W band that pass certain ellipse fitting criteria. \citet{guo_2024} identify bars using both ellipse fitting techniques and visual inspection on the F200W and F444W bands separately and combined. In order to make the most fair comparison to our work, the bar fractions shown in Figure \ref{fig:main} are the bar fractions determined by visual inspection of bars found in either the F200W or F444W band (column 7 in their Table 3). However, \citet{guo_2024} do not distinguish between weak and strong in their bar fractions. The effects of identifying bars on single-band images compared to multi-band coloured images is discussed in more detail in Section \ref{sec:disc_single_multi_band}. Lastly, there are also some differences in the sample selection, which are noted in greater detail in Section \ref{sec:disc_comparing}. All of these differences make a direct comparison more difficult. Nevertheless, our corrected bar fractions are well within the errors of the bar fractions found in \citet{leconte_2024}. This is also the case for the bar fraction of \citet{guo_2024}, except in the $1.5 < z < 2.0$ bin, where our bar fractions are higher. However, they do not correct for redshift effects, while we do. Additionally, their bar fraction in that bin based on ellipse fits, rather than visual classification, equals $11.8\pm2.3\%$, which is consistent with our corrected bar fraction after taking the uncertainties into account.

One of the more interesting results is that the bar fraction is still relatively high at high redshifts ($2.0 < z < 2.5$), with the raw and corrected bar fractions equalling \todo{$17^{+6}_{-4}$\%} and \todo{$16^{+6}_{-4}$\%}, respectively. This is in contrast with some simulations that predict that very few bars should exist beyond $z = 1 - 1.5$ \citep{kraljic_2012, algorry_2017, reddish_2022}. This pushes back the so-called `epoch of bar formation' from $z = 0.7 - 1$ \citep{kraljic_2012, melvin_2014, simmons_2014,donohue_keyes_2019} to at least $z = 2.5$. This is discussed in more detail in Section \ref{sec:disc_epoch_bar_formation}. Simulations predict that the bar formation timescale varies from anywhere between 1 to 5 Gyr \citep{saha_2013, algorry_2017}. \citet{bland_hawthorn_2023,bland_hawthorn_2024} even find bar formation timescales of $\sim 500$ Myr for baryon-dominated discs. Our results are most consistent with a bar formation timescale on the lower end of that range, as we still find a significant number of bars at $z = 2.5$.

Finally, the bar fractions are low in our highest redshift bin, $3.0 < z < 4.0$,  equalling \todo{$6^{+7}_{-2}$\%} and \todo{$3^{+6}_{-1}$\%} for the raw and corrected bar fractions, respectively. This suggests that the stellar discs at these redshifts are still dynamically hot or have not yet had sufficient time to form bars, although various observational effects might affect our bar fraction at these higher redshifts. This is discussed in more detail in Section \ref{sec:disc_decreasing_bar_fraction}. Also note that the sample size is relatively low in this bin, with only \todo{33} featured galaxies and a total of \todo{2} identified bars. Increasing the sample size with morphological classifications from other JWST imaging surveys, such as COSMOS-Web \citep{casey_2023} and the \glswithcite{JADES}{\citealt{eisenstein_2023}}, would help significantly to provide better constraints on the bar fraction at these higher redshifts.

\subsection{Epoch of bar formation and dynamically cold discs}
\label{sec:disc_epoch_bar_formation}

Previous studies have suggested that the `epoch of bar formation' begins around $z \sim 0.7 - 1$ \citep{kraljic_2012, melvin_2014,simmons_2014,donohue_keyes_2019}. This would be when galaxies started becoming disc dominated and dynamically cool. This is supported by simulations that find little to no bars at redshifts higher than $z = 1.0 - 1.5$ \citep{kraljic_2012, algorry_2017, reddish_2022}. However, the results presented in this work suggest that a considerable amount of bar formation might occur at higher redshifts than previously thought, as we still find a significant number of bars up to $ z \approx 2.5$. While the corrected bar fractions are clearly decreasing with redshift, the decrease is only modest between $0.5 < z < 2.5$, going from \todo{$25^{+6}_{-4}$}\% in the lowest redshift bin to \todo{$16^{+6}_{-4}$}\% at $2.0 < z < 2.5$. Instead, the biggest change in the corrected bar fraction happens when we move to the next bin, $2.5 < z < 3.0$, where it drops to \todo{$8^{+6}_{-2}$}\%. The raw bar fractions show a more gradual decrease. This implies that galaxies at these higher redshifts were dynamically cool enough to host bars. Alternatively, these galaxies might be dynamically hot, but baryon-dominated, since \citet{bland_hawthorn_2023,bland_hawthorn_2024} have recently shown that bars can form in dynamically hot galaxies, depending on ratio of baryons to dark matter of the disc. Either way, we find a significant number of bars up to $ z \approx 2.5$, which suggests that bar-driven secular processes could be important at higher redshifts than previously expected. Similar conclusions are also drawn from recent simulations, which show that bars can exist at high redshifts ($z \geq 2$, \citealt{rosas_guevara_2022, bi_2022, zana_2022, fragkoudi_2020, fragkoudi_2025}). \citet{leconte_2024} also come to a similar conclusion, as they also find a significant number of bars at $z > 2$ and note that this implies bar-driven evolution at these higher redshifts. 

Our corrected bar fractions therefore seem to suggest that stellar discs at $z > 2.5$ are not able to form and sustain bars as efficiently, as they might be too dynamically hot, thick and turbulent \citep{kraljic_2012,sheth_2012,aumer_2017,ghosh_2023}. The corrected bar fraction is very low in the highest redshift bin ($3 < z < 4$), equalling \todo{$3^{+6}_{-1}$}\%. It would be interesting to see if we can find a significant population of bars and dynamically cold stellar discs at $z > 4$, though some barred galaxies have already been found at these redshifts. For example, \citet{smail_2023} found a galaxy at $z = 4.26$ with a linear bar feature, \citet{tsukui_2024} found a barred galaxy at $z = 4.4$, and \citet{neeleman_2023} found a thin rotating disc at $z = 6.5$. Finally, the highest redshift featured disc in \citet{smethurst_2025} was found at $z = 5.5$. Nevertheless, at the moment, these galaxies seem more rare. These results do depend on correctly adjusting for any redshift effects, which is discussed in greater detail in Section \ref{sec:disc_decreasing_bar_fraction}.

\subsection{Disentangling bar strength over redshift}
\label{sec:disc_bar_strength}

The evolution of the strong bar fraction and the weak bar fraction over redshift is shown in Figure \ref{fig:bar_strength}. The weak bar fraction decreases with redshift and is consistent within the error bars with the observations of \citet{leconte_2024}. Interestingly, the strong bar fraction remains constant between $0.5 < z < 2.5$. Table 2 in \citet{leconte_2024} also shows a similar result. They find 20 weak bars (corresponding to 12.7\%) between $1 < z < 2$, but only 5 (6.9\%) in $2 < z < 3$, a significant decrease. Meanwhile, the decrease in the number of strong bars is more moderate: they find 8 strong bars (corresponding to 5.1\%) in $1 < z < 2$ and 5 (6.9\%) in $2 < z < 3$. Multiple observations and simulations allow for the possibility that bars are long-lived robust structures \citep{jogee_2004,debattista_2006,athanassoula_2013,gadotti_2015, perez_2017,ghosh_2023,fragkoudi_2025,lopez_2024}, although other studies suggest that they might be short-lived transient features and can get destroyed or weakened \citep{bournaud_2002,shen_2004,athanassoula_2005b, bournaud_2005,ghosh_2021}. Our results support the idea that the strong bars observed in this work are long lived structures that were formed around $z \approx 2 - 2.5$. This is consistent with other observations. For example, \citet{perez_2017} found that the bar in NGC 6032 was formed around 10 Gyr ago, which corresponds to $z \sim 1.86$. Similarly, the bar in NGC 4371 was estimated to have been formed at $z \sim 1.8$ \citep{gadotti_2015}. Another way to explain the constant bar fraction is that the rate of strong bar formation is roughly similar to the rate of strong bar destruction. \citet{simmons_2014} note that a combination of increased merger rates, high gas fractions and dynamically warm discs could lead to the formation of short-lived bar structures at higher redshifts. This could result in a low, but stable bar fraction. \citet{smethurst_2025} make a similar argument for the existence of such a high number of featureless discs at these redshifts. 

Interestingly, the strong bar fraction in the lowest redshift bin in this work is significantly lower than the bar fractions found in other work that study the local Universe. We find a strong bar fraction of \todo{$4^{ +4 }_{ -1 }\%$} at $0.5 < z < 1.0$, while \citet{geron_2021} find a strong bar fraction of 15.5\% at $0.01 < z < 0.05$. This implies that there is a significant evolution in the strong bar fraction between $z = 0.05$ and $0.5 < z < 1.0$. This is supported by \citet{melvin_2014}, who observed a significant change in the strong bar fraction between $0.4 < z < 1$. If we assume that strong bars are robust features that cannot be destroyed, then these findings imply that there are two bursts of strong bar formation. Our results suggest that a first burst happened around $z = 2 - 2.5$, which corresponds to a lookback time of 10.2 - 10.9 Gyr. This period brought the strong bar fraction from $\sim\todo{0}\%$ to $\sim\todo{4}\%$. No new strong bars are formed and the strong bar fraction remains constant until a second, more recent, burst of strong bar formation happened at $z < 1$. This is consistent with the epoch of bar formation around $z = 0.7 - 1$ mentioned by other work \citep{kraljic_2012,melvin_2014,simmons_2014}, but one that would specifically produce strong bars. This corresponds to a lookback time of 6.3 - 7.7 Gyr. This brings the strong bar fraction from $\sim\todo{4}\%$ to $\sim\todo{15}\%$ in the local Universe. It would be interesting to look at the properties of the host galaxies with strong bars that formed around $z = 0.7 - 1$ and the host galaxies with strong bars that formed around $z = 2 - 2.5$ and see if there are any obvious differences.

The weak bar fraction reported by \citet{geron_2021} is 28.1\% at $0.01 < z < 0.05$, which is almost within the error bars of the corrected weak bar fraction found in the lowest redshift bin here, \todo{$21^{+6}_{-4}\%$}. This suggests that a moderate change in the weak bar fraction is expected between $0.05 < z < 0.5$, but not nearly as dramatic as the one for strong bars.

Bars tend to grow longer and stronger over time \citep{lynden_bell_1972,sellwood_1981,athanassoula_2003,martinez_valpuesta_2006,athanassoula_2013, algorry_2017}. A growing weak bar fraction with decreasing redshift and a constant strong bar fraction can be explained by assuming that the bar strengthening timescale is larger than the bar formation timescale. Weak bars are continuously formed, which causes the weak bar fraction to increase. However, they need more time before they become strong bars, which keeps the strong bar fraction constant. This is consistent with the simulations of \citet{algorry_2017}, who find that even the fastest growing bars in their sample take $\sim$4-5 Gyr to fully form. As mentioned in Section \ref{sec:disc_barfrac}, simulations predict bar formation timescales between 500 Myr - 5 Gyr \citep{saha_2013, algorry_2017,bland_hawthorn_2023,bland_hawthorn_2024}, though our results are most consistent with the values on the lower end of that scale. An alternative interpretation is that another factor is preventing weak bars from becoming strong until $z \sim 0.7-1$.

It is likely that observational biases affect the interpretation of these results. For example, as weaker bars are shorter, they will be easier to miss at higher redshifts, which can explain the observed decrease in weak bar fraction over redshift. It is also possible that it is harder to accurately classify bar strength at higher redshifts. Naively, you could expect that a strong bar at higher redshift could be misclassified as a weak bar, because it would be harder to see and appear smaller. However, previous work has shown that volunteers likely classify bars into strong and weak based on the relative size of the bar compared to the disc \citep{geron_2021}. As discs are typically less bright than bars, this ratio would increase with redshift. Perhaps the opposite is true: that weak bars get misclassified as strong bars at higher redshift.

There is also the added complexity that, as mentioned above, bars tend to grow longer and stronger over time \citep{lynden_bell_1972,sellwood_1981,athanassoula_2003,martinez_valpuesta_2006,athanassoula_2013, algorry_2017}. This implies that the strong bar fraction should decrease with redshift, as we expect more stronger bars in the low redshift Universe. However, the observational bias described above could prevent us from observing this trend. Additionally, as we expect bar length and strength to change over time, perhaps what we consider ``strong'' and ``weak'' changes at higher redshifts as well. Finally, the redshift correction applied in this work could be different for weak and strong bars as well, though this lies beyond the scope of this work. A study looking at detection thresholds for different bar lengths and bar strengths at these higher redshifts will be paramount to help clarify this issue.

While the strong bar fraction is roughly constant between $0.5 < z < 2.5$, it drops significantly at higher redshift. It is likely that bars have not had sufficient time to become strong before $z \sim 2.5$. We only find \todo{one} strong bar between $2.5 < z < 3$, which corresponds to a raw strong bar fraction of \todo{2\%}, and we find \todo{no} strong bars at the highest redshift bin ($3 < z < 4$). This is interesting, as we still identify a small number of weak bars (\todo{3} and \todo{2} in each bin, respectively), and presumably strong bars would be easier to spot at these higher redshifts than weak bars. However, redshift effects at these large distances might make strong bars look like weak bars. It is also important to keep in mind that the small sample sizes at the high redshift bins make drawing strong conclusions difficult.

\subsection{Bar formation mechanisms at high redshifts}
\label{sec:disc_bar_formation}

Simulations show that bars can spontaneously form through instabilities in rotationally supported cold galactic discs \citep{hohl_1971, ostriker_1973,efstathiou_1982, sellwood_1993}. These bars start out weaker and grow longer and stronger over time \citep{lynden_bell_1972,sellwood_1981,athanassoula_2003,martinez_valpuesta_2006,athanassoula_2013, algorry_2017}. However, due to the higher rate of interactions and mergers at higher redshifts \citep{lotz_2011,duan_2025} and discs being too dynamically hot \citep{kraljic_2012, reddish_2022}, bar formation through disc instabilities might be rarer at higher redshifts. Mergers and tidal interactions can also directly trigger bar formation \citep{noguchi_1987,elmegreen_1991,lang_2014,peschken_2019,merrow_2024}, although it depends on the orbital configuration of the encounter \citep{gajda_2017,lokas_2018}. This suggests that bars at higher redshifts might preferentially be formed through tidal interactions, rather than through disc instabilities. This is consistent with the preliminary results of \citet{guo_2024}, who found that a larger fraction of $z > 1.5$ barred galaxies have nearby neighbours or showed signs of a tidal interaction, compared to unbarred galaxies.

Interestingly, simulations show that bars formed by mergers or tidal interactions tend to be stronger compared to bars formed through disc instabilities \citep{peschken_2019,fragkoudi_2025}. This can help to explain the strong and weak bar fraction trends we see in Figure \ref{fig:bar_strength}. A decreasing weak bar fraction with redshift can be explained by disc instabilities, which form initially weaker bars, becoming increasingly rare at higher redshifts. Similarly, the flattening of the strong bar fraction can be interpreted as fewer bars being formed through tidal interactions at lower redshifts. Thus, while other interpretations are possible, our results are consistent with disc instabilities being the dominant mode of bar formation at lower redshifts, while bar formation through interactions and mergers is more common at higher redshifts.

\subsection{Single-band images or multi-band images?}
\label{sec:disc_single_multi_band}

\begin{figure}
	\includegraphics[width=\columnwidth]{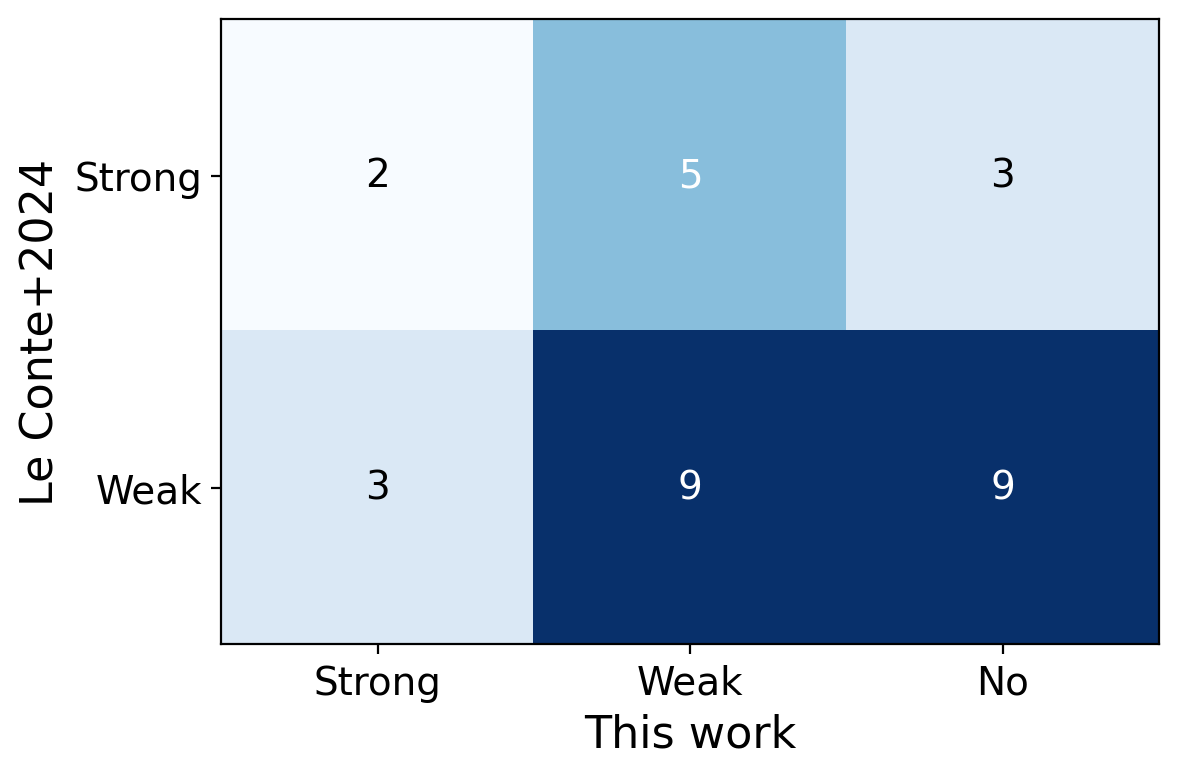}
    \caption{Confusion matrix of the 31 galaxies that overlap between this work and \citet{leconte_2024}. This shows that there is some disagreement on bar type classification, which is likely due to the fact that we use coloured multi-band images to identify bars, while \citet{leconte_2024} use single-band images.}
    \label{fig:leconte_cm}
\end{figure}

As mentioned above, volunteers are shown multi-band coloured images in GZ CEERS, while most other studies typically use single-band images when identifying bars (e.g. see \citealt{liang_2024,leconte_2024}). Interestingly, \citet{guo_2024} show that, in every redshift bin, some bars are only identified in either the F200W or F444W band, but not in the other. This is likely because shorter wavelength bands have smaller \gls{FWHM}, allowing them to detect smaller structures, while longer wavelength bands are less affected by dust and star formation. In theory, combining the different bands into coloured images allows us to make use of the advantages (and disadvantages) of all bands at the same time. We test this statement by cross-matching our parent sample to the barred galaxies found by \citet{leconte_2024}, who identified bars in JWST \gls{CEERS} using a combination of visual inspection and ellipse fitting techniques in the F444W band. We find that \todo{31} of the 39 barred galaxies found in \citet{leconte_2024} are also in our parent sample. However, interestingly, not all of their bars are classified as barred in our study. A confusion matrix of our classifications compared to their classifications is shown in Figure \ref{fig:leconte_cm}, which shows that \todo{12} of the \todo{31} galaxies they classify as barred, are classified as unbarred here. When we split up by bar strength, we find that we classify \todo{70}\% of their strong bars as barred and \todo{57}\% of their weak bars as barred. A likely reason for this discrepancy is the difference in bands used to identify bars.

\begin{figure*}
	\includegraphics[width=\textwidth]{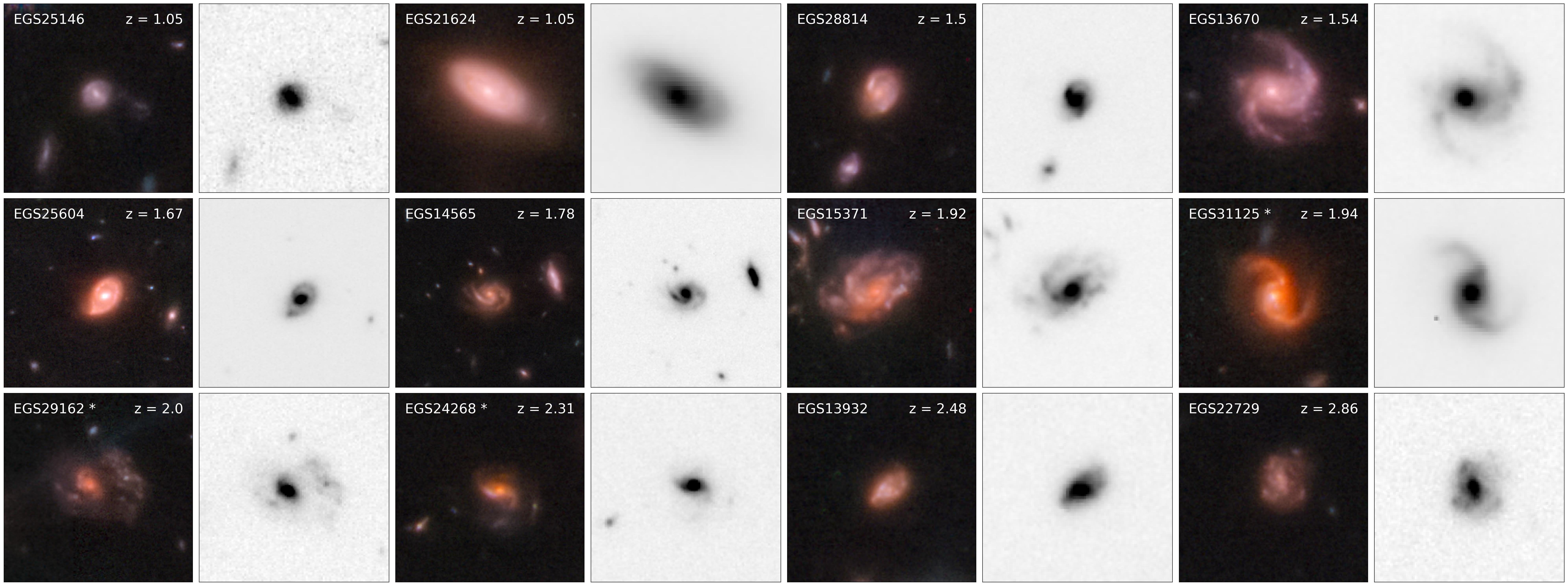}
    \caption{The 12 galaxies that are classified by \citet{leconte_2024} as either weakly or strongly barred, while they are classified as unbarred by GZ CEERS.  The CANDELS ID of each galaxy is shown in the top left corner of each image, while the redshift is shown in the top right corner. The bars with a `*' next to their CANDELS ID are classified by \citet{leconte_2024} as strongly barred, while the rest are weakly barred. For every galaxy, we show both the coloured GZ CEERS image and the single-band F444W image in greyscale. Note that the grey colour scale is not identical to the one used in \citet{leconte_2024}, though we tried to approximate it as closely as we could.}
    \label{fig:leconte_comparison}
\end{figure*}

Both the colour GZ CEERS images and the single-band F444W images of the \todo{12} galaxies that \citet{leconte_2024} classify as barred but we classify as unbarred are shown in Figure \ref{fig:leconte_comparison}. The coloured images of these galaxies show no clear evidence of bars. Interestingly, the F444W images do reveal elliptical structures in the centres of some of these galaxies, which could potentially be bars. The high \gls{FWHM} of this band makes discerning more detail difficult. At the same time, it is also likely that these structures are obscured by dust in the other filters, as they are not visible in the coloured images. This figure clearly illustrates the problem with which we are faced, and it is not clear whether using single bands or multiple bands is the better approach. 

Additionally, among galaxies that we both classify as barred, we do not always agree on the strength of the bars. This is likely because the method of classifying bar strength differs between both studies. We classify bars as strong or weak as outlined in Section \ref{sec:sample_selection}. In \citet{leconte_2024}, five authors classified galaxies as `barred', `maybe-barred' or `unbarred'. They said a galaxy is strongly barred if at least three out of five experts voted for barred. A galaxy is weakly barred if two out of five experts voted for barred or at least three out of five voted for maybe-barred. This is a reasonable approach, but there is a small nuance here that can manifest itself in differences in classification. This approach suggests that weak bars are bars that you are more unsure about. However, it is possible to be sure that a galaxy has a bar that is weak; though such a bar would be classified as a strong bar with this approach. As an illustration, in our sample, there are \todo{24} weak bars that have $p_{\textrm{weak bar}} > 0.6$, which means that volunteers are convinced that there is a bar in these galaxies, and they are confident that these bars are weak. Both approaches to bar strength classification have merit, but this difference can explain some of the discrepancies observed here. 

While identifying bars is easy in principle, in practise there are decisions to be made that will affect the final outcome. It is clear that different methods find slightly different things. We want to be clear that both approaches are valid, but it is good to be aware of these potential differences. Interestingly, despite our slightly different methods, our overall bar fractions are still consistent with each other (see Figure \ref{fig:main}). So while we might disagree on individual bar classifications, we agree on the population level.

\subsection{Difficulties with comparing bar fractions}
\label{sec:disc_comparing}

Figure \ref{fig:main} shows the raw and corrected bar fractions found in this work, as well as bar fractions from multiple other observational studies. However, it is not obvious that we can even directly compare bar fractions between different studies. This is because the bar fraction is very heavily affected by properties of the sample (e.g. stellar mass and SFR; \citealp{barazza_2008,sheth_2008, cameron_2010, nair_2010b, masters_2011,cheung_2013, melvin_2014, gavazzi_2015, fraser_mckelvie_2020b, geron_2021}), which can differ dramatically between studies. For example, we create a volume-limited sample to properly compare bar fractions across redshift bins. To achieve a similar goal, \citet{guo_2024} only considered galaxies with stellar mass $M_{*} > 10^{10} M_{\odot}$. We have a significant population of galaxies with stellar masses lower than that, which makes a direct comparison complicated. \citet{leconte_2024} creates a mass-complete sample based on the 95\% empirical completeness found in \citet{duncan_2019}. This means that their sample includes lower mass galaxies and is more comparable to ours, though  some differences are inevitable. These differences in sample selection are also likely why our results are in better agreement with \citet{leconte_2024} than with \citet{guo_2024}. 

As noted above, it is also known that the wavelength used to identify bars strongly affects the final bar fractions. Bar fractions in optical wavelengths in low redshift studies are typically 43\%-52\% \citep{marinova_2007,barazza_2008,aguerri_2009,buta_2019,geron_2021}, while they rise to 59-73\% when using infrared wavelengths \citep{eskridge_2000,marinova_2007,menendez_delmestre_2007,sheth_2008}, as infrared wavelengths are less affected by star formation and dust \citep{erwin_2018}. In contrast, bands with smaller \gls{FWHM} are able to detect smaller and weaker bars, further complicating this issue. Additionally, multiple methods to find and identify bars exist, ranging from visual classifications (e.g. see \citealt{nair_2010, simmons_2014}) to automated ellipse fitting techniques (e.g. see \citealt{jogee_2004,marinova_2007}). However, these techniques do not always agree. For example, \citet{lee_2019} show that ellipse fitting techniques miss $\sim$15\% of strong bars compared to visual classification. \citet{guo_2024} also find a slightly different number of bars depending on whether they use ellipse fitting techniques or visually identify bars, though the two methods are largely consistent with each other (see their Table 3).

Correctly estimating the number of galaxies with bars (i.e. the numerator of Equation \ref{eq:barfrac}) is crucial to correctly estimate the bar fraction. However, it is almost equally important to correctly estimate the number of galaxies with discs (i.e. the denominator of Equation \ref{eq:barfrac}). Furthermore, different studies will use slightly different denominators (e.g. galaxies with discs, not edge-on discs or simply featured galaxies), which can change the final bar fraction. For example, the inclusion of featureless discs in the denominator (see Section \ref{sec:corrections_fd}) significantly altered the final values of the bar fractions found in this work. 

These issues are not limited to observations. Comparing bar fractions can also be challenging between simulations, as the sample selection can vary significantly between different studies. For example, the simulations of \citet{rosas_guevara_2022} look at galaxies with $M_{*} > 10^{10} M_{\odot}$ in a large cosmological volume in TNG50. \citet{fragkoudi_2025} focusses on 39 progenitors of Milky Way-like galaxies in their Auriga simulations with present-day stellar masses $M_{*} > 10^{10} M_{\odot}$. Finally, \citet{kraljic_2012} look at 33 galaxies with present-day stellar masses between $10^{10} - 10^{11} M_{\odot}$. It is also important to emphasize that the same galaxies are traced back in time in studies using zoom-in simulations (such as \citealt{fragkoudi_2025,kraljic_2012}). Therefore, especially at higher redshifts (e.g. $z > 2-3$), the distribution of stellar masses from studies using zoom-in simulations and those using large cosmological volumes (e.g. \citealt{rosas_guevara_2022}) can be very different. As the zoom-in simulations specifically trace Milky Way progenitors, they will not contain galaxies with very high stellar masses (e.g. $M_{*} > 10^{10} M_{\odot}$) at these higher redshifts. In contrast, such high mass galaxies could exist at high redshifts in simulations using large cosmological volumes, depending on the details of the simulation. Thus, as with observations, these differences in sample selection will inevitably result in differences in the bar fraction found in simulations. In order to make a precise comparison between simulations and observations, the same selection function needs to be used for both simulations and observations.

All of these effects make comparing bar fractions between studies difficult. This becomes especially apparent when you look at bar fractions found at lower redshifts, shown in Figure \ref{fig:main}, which can vary between 20\% to 70\%. While it is still useful to compare bar fractions and their trends between studies, it is important to keep these issues in mind.

\subsection{Is the bar fraction truly decreasing?}
\label{sec:disc_decreasing_bar_fraction}

We discussed in Section \ref{sec:disc_barfrac} how the bar fraction decreases with redshift and what this implies for the epoch of bar formation in Section \ref{sec:disc_epoch_bar_formation}. However, how confident are we that we corrected for all possible observation biases and truly detected all bars in the sample? 

Detecting bars becomes more difficult at higher redshifts because of a combination of resolution and cosmological dimming effects. We tried to account for these in our corrected bar fractions. As explained in Section \ref{sec:corrections_z}, applying the appropriate redshift correction is not straightforward. The redshift correction used in this work is based on the single band simulations of \citet{liang_2024}. However, we use multi-band coloured images, and it is not obvious which of the single band simulations would be most applicable to our multi-band images. We therefore opted for the most minimal redshift correction, which implies that our corrected bar fractions should be interpreted as lower limits. The true bar fraction is likely higher and the true bar fraction evolution over redshift is probably less extreme than detected. 

The final bar fractions heavily depend on which correction is used. To illustrate this point, Figure \ref{fig:z_correction} shows that if we were to use the redshift correction based on the F345W band, our corrected bar fractions would be consistent with being constant. \citet{liang_2024} noted a similar finding based on the observations of \citet{leconte_2024}. Other studies have also suggested constant bar fractions. The simulations of \citet{rosas_guevara_2022} find that the bar fraction remains almost constant between 0.5 $\geq$ $z$ $\geq$ 3.0, which manifests itself as a decreasing bar fraction when the limited spatial resolution of observations is taken into account (though they find a decline in bar fraction beyond $z$ = 4.0). Additionally, the observations of \citet{elmegreen_2004} and \citet{jogee_2004} also support a constant bar fraction between $0.0 < z \lessapprox 1.1$. It is unlikely that the correction based on the F345W band would be the appropriate one to use for our multi-band images, but this finding is still worth considering. 

Another complication arises from the large redshift range that we are probing, as the rest-frame wavelength of the galaxies changes significantly. For example, the F200W band probes the \gls{NIR} (the pivot wavelength of F200W equals $\sim1.988\mu$m\footnote{\url{https://jwst-docs.stsci.edu/jwst-near-infrared-camera/nircam-instrumentation/nircam-filters}}). However, the corresponding rest-frame wavelength will go from $\sim1.3 \mu$m at $z=0.5$ to $\sim400$nm at $z = 4$, which is in the blue part of the optical spectrum. This is a problem, as we know that we observe more bars in the \gls{NIR} compared to optical wavelengths \citep{erwin_2018} and implies that a decreasing bar fraction is expected from this effect alone, even if the true underlying bar fraction would remain constant. 

A final issue that further complicates this problem is that simulations show that bars tend to become longer and stronger over time \citep{sellwood_1981,athanassoula_2003,martinez_valpuesta_2006,athanassoula_2013,algorry_2017}. This means that bars will be even harder to find at higher redshifts, as they will be shorter and the detection threshold of L$_{\rm bar} / FWHM > 2$ (e.g. see \citealt{erwin_2018, liang_2024, xu_2024}) will be less likely to be met. In summary, it is difficult to disentangle whether the bar fraction is truly decreasing, or if this is simply the result of observational biases caused by redshift. This is true across all redshift ranges, although trends at higher redshifts (e.g. 2.0 - 4.0) are more uncertain than those at lower redshifts (e.g. 0.0 - 2.0). More detailed simulations looking into expected bar fractions for multi-band coloured images over a wide redshift range, while considering both strong and weak bars, will help to obtain a more accurate redshift correction and to better constrain these issues.

\section{Conclusions}
\label{sec:conclusions}

Studying the bar fraction at high redshifts is crucial as it helps uncover when galaxies become dynamically cool enough to host bars and how important the role of bar-driven secular processes is at these high redshifts. The bar fraction can also help to probe the lifetime and robustness of bars. In this work, we have identified strong and weak bars over a wide range of redshifts ($0.5 < z < 4.0$) using volunteer classifications on multi-band coloured images obtained from \gls{CEERS}. The use of multiple bands allows us to take advantage of multiple wavelengths. The full morphological catalogue, GZ CEERS, is described in greater detail in \inprep{Masters et al. (in prep)}. There are \todo{668} galaxies in our parent sample, of which \todo{161} are barred galaxies. Similarly, we found that \todo{87} out of the \todo{398} galaxies in the volume-limited subsample host bars. This is the largest number of barred galaxies found at these higher redshifts to date. We also differentiate between strongly and weakly barred galaxies, correct for redshift effects and are careful to include featureless discs. We applied the most minimal realistic corrections, which means that our corrected bar fractions should be interpreted as lower limits and that the true bar fraction likely lies above the quoted values.

We also want to highlight that, while in principle the notion of a bar fraction seems relatively simple, in practise it is difficult to obtain. It depends on the exact definition used (see Equations \ref{eq:barfrac}, \ref{eq:barfrac_gz}, \ref{eq:barfrac_f23} and \ref{eq:barfrac_corr}), which bands are used, and what kind of sample you consider. The bar fraction is also very sensitive to any corrections applied. There are a lot of complicated observational biases that are hard to disentangle. Additionally, the sample size quickly becomes low at higher redshifts. Obtaining reliable bar classifications for JWST surveys other than \gls{CEERS}, such as COSMOS-Web \citep{casey_2023} and the \glswithcite{JADES}{\citealt{eisenstein_2023}}, will be crucial to better constrain the bar fraction at these high redshifts and to possibly probe even higher redshifts.

Our main findings are that:

\begin{itemize}
    \item In agreement with other work, we find that the overall bar fraction decreases over redshift with $>$$3\sigma$ significance. The raw bar fraction equals \todo{$39^{+6}_{-6}$\%} at $0.5 < z < 1.0$ and goes down to \todo{$6^{+7}_{-2}$\%} at $3.0 < z < 4.0$. Similarly, the corrected bar fraction starts at \todo{$25^{+6}_{-4}$\%} at $0.5 < z < 1.0$ and decreases to \todo{$3^{+6}_{-1}$\%} at $3.0 < z < 4.0$. 

    \item Despite a decreasing bar fraction, we still find a significant number of bars at $z = 2.5$. The biggest decrease in bar fraction happens between $2.0 < z < 2.5$ and $2.5 < z < 3.0$, where the corrected bar fraction goes from \todo{$16^{+6}_{-4}$}\% to \todo{$8^{+6}_{-2}$\%}, respectively. This suggests that discs are dynamically cool enough to host a significant number of bars at these high redshifts or that they are baryon-dominated. Additionally, this suggests that bar-driven secular evolution could have been playing a crucial role in the evolution of galaxies since $z = 2.5$. 
    
    \item We find that the strong bar fraction is constant around \todo{4\%} between $0.5 < z < 2.5$. This is consistent with the idea that strong bars are robust long-lived structures, unless the rate of bar formation is similar to the rate of bar destruction. Additionally, since the bar fraction in the present-day Universe equals $\sim15\%$, a significant evolution in the strong bar fraction is expected between $0.0 < z < 1.0$. The weak bar fraction consistently goes down with redshift, which suggests that weak bars are formed anywhere between $0.5 < z < 2.5$. These results are consistent with disc instabilities (which tend to form weaker bars that become stronger over time) being the dominant mode of bar formation at lower redshifts, while bar formation through interactions and mergers (which tend to form stronger bars) is more common at higher redshifts. 

\end{itemize}



\section*{Acknowledgements}
The data in this paper are the result of the efforts of the Galaxy Zoo volunteers, without whom none of this work would be possible.

This research was supported by the International
Space Science Institute (ISSI) in Bern, through ISSI International Team project 23-584 (Development of Galaxy Zoo: JWST). 

TG is a Canadian Rubin Fellow at the Dunlap Institute. MW is a Dunlap Fellow at the Dunlap Institute. The Dunlap Institute is funded through an endowment established by the David Dunlap family and the University of Toronto. ILG has received the support from the Czech Science Foundation Junior Star grant no. GM24-10599M. BDS acknowledges support through a UK Research and Innovation Future Leaders Fellowship [grant number MR/T044136/1]. RJS gratefully acknowledges support through the Royal Astronomical Society Research Fellowship. RC thanks support by the Natural Sciences and Engineering Research Council of Canada (NSERC), [funding reference \#DIS-2022-568580]. LFF, HR and KBM acknowledge partial support from the US National Science Foundation under grants IIS 2006894, IIS 2334033 and NASA award \#80NSSC20M0057. FF is supported by a UKRI Future Leaders Fellowship (grant no. MR/X033740/1). MRD acknowledges support from the NSERC through grant RGPIN-2019-06186, the Canada Research Chairs Program, and the Dunlap Institute at the University of Toronto. The project that gave rise to these results received the support of a fellowship from the “la Caixa” Foundation (ID 100010434). The fellowship code is LCF/BQ/PR24/12050015. LC acknowledges support from grants PID2022-139567NB-I00 and PIB2021-127718NB-I00 funded by the Spanish Ministry of Science and Innovation/State Agency of Research  MCIN/AEI/10.13039/501100011033 and by “ERDF A way of making Europe”. 

This work made use of Astropy:\footnote{http://www.astropy.org} a community-developed core Python package and an ecosystem of tools and resources for astronomy \citep{astropy_2013, astropy_2018, astropy_2022}.

This work is based on observations taken by the CANDELS Multi-Cycle Treasury Program with the NASA/ESA HST, which is operated by the Association of Universities for Research in Astronomy, Inc., under NASA contract NAS5-26555.


%






\appendix

\section{Full list of barred galaxies found in GZ CEERS}
\label{app:all_images}

\begin{figure*}
	\includegraphics[width=\textwidth]{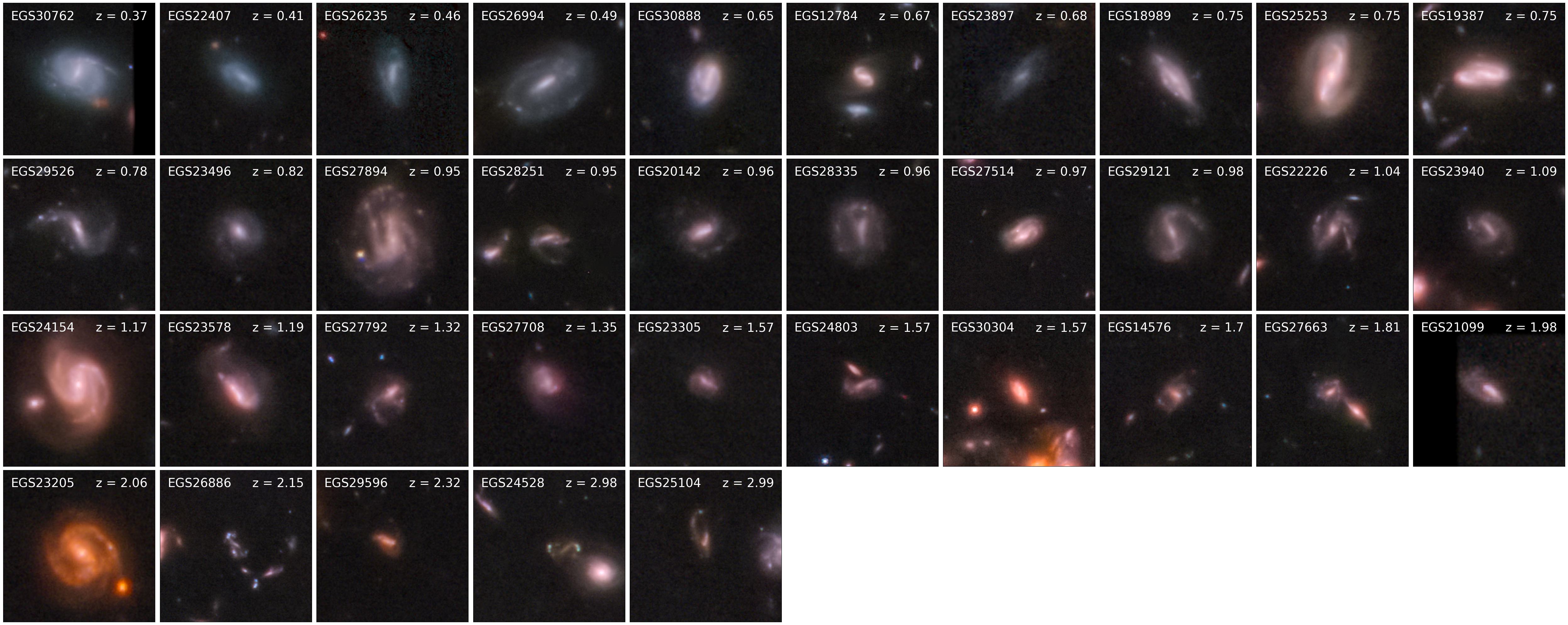}
    \caption{Colour images of all strongly barred galaxies found by volunteers in the parent sample of GZ CEERS. The colouring matches how the images were presented to the volunteers. The CANDELS ID of each galaxy is shown in the top left corner of each image, while the redshift is shown in the top right corner.}
    \label{fig:all_sb}
\end{figure*}

\begin{figure*}
	\includegraphics[width=\textwidth]{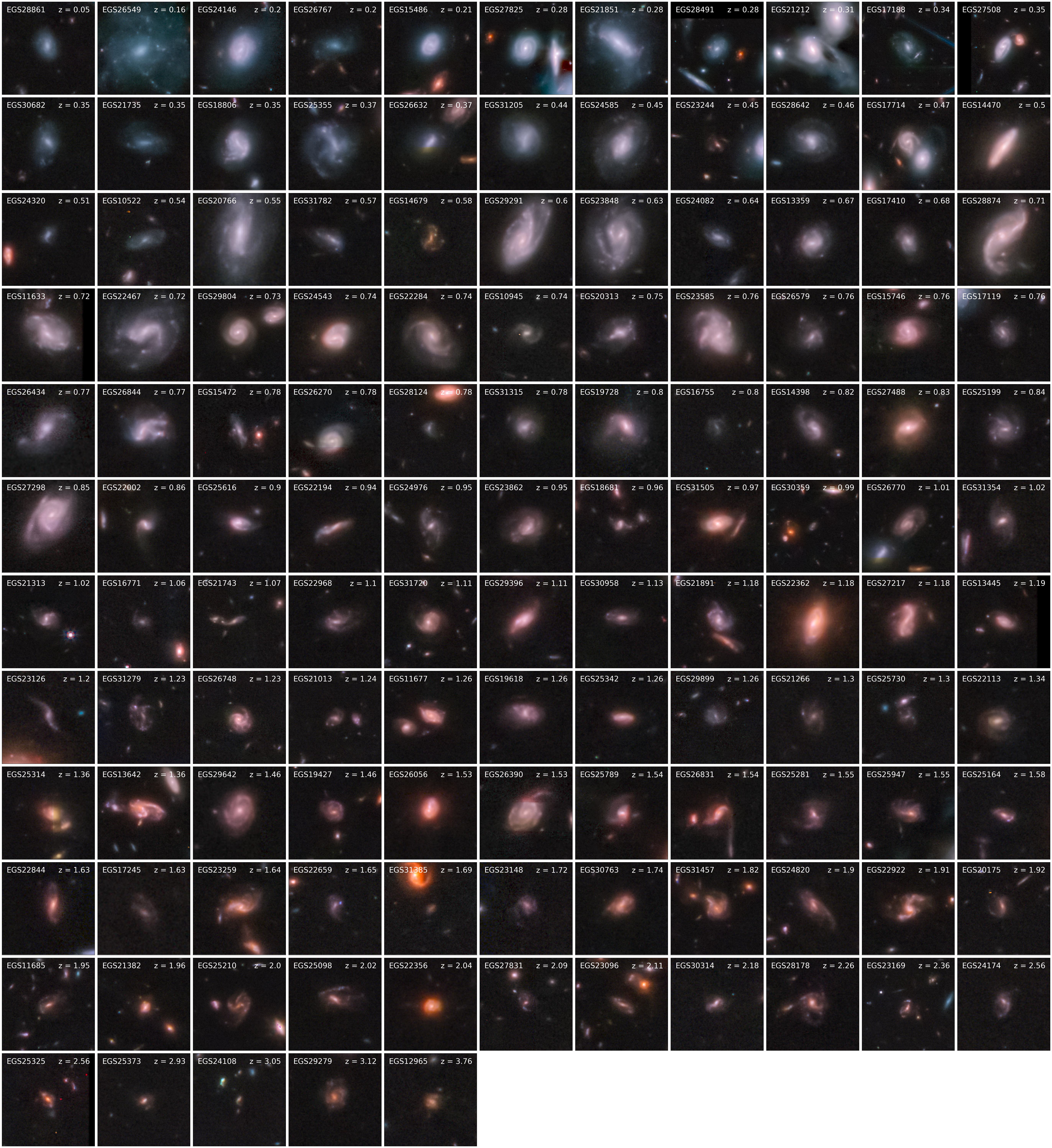}
    \caption{Colour images of all weakly barred galaxies found by volunteers in the parent sample of GZ CEERS. The colouring matches how the images were presented to the volunteers.  The CANDELS ID of each galaxy is shown in the top left corner of each image, while the redshift is shown in the top right corner.}
    \label{fig:all_wb}
\end{figure*}

Figure \ref{fig:overview} showed examples of strongly and weakly barred galaxies found in GZ CEERS. Here, we show coloured CEERS cutouts of the \todo{35} strongly barred (Figure \ref{fig:all_sb}) and the \todo{126} weakly barred (Figure \ref{fig:all_wb}) galaxies found in the parent sample. The full sample, complete with coordinates, redshifts, identifiers and bar strength classifications, can be found in machine-readable format \href{https://github.com/tobiasgeron/gz_ceers_bar_fractions}{here}\footnote{\url{https://github.com/tobiasgeron/gz_ceers_bar_fractions}}.

\section{Effect of resolution on bar identification}
\label{app:resolution}

\begin{figure*}
	\includegraphics[width=\textwidth]{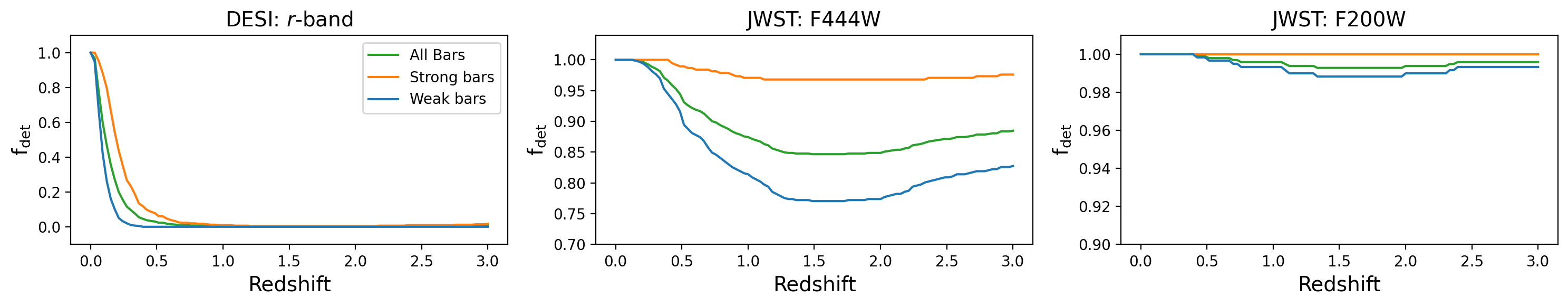}
    \caption{The detection fraction (f$_{\rm det}$) of the GZ DESI bar sample over redshift for the DESI $r$-band (left panel), the JWST F444W band (middle panel) and the JWST F200W band (right panel). Note the different scales on the y-axis for the three panels. We show the detection fraction for strong bars (orange), weak bars (blue) and all bars (green). This figure shows us that resolution effects are not the limiting factor for detecting bars in JWST images, as the fraction of detected bars in the shorter wavelength bands (e.g. F200W) is $\sim$99\%, even for the shorter, weaker bars.}
    \label{fig:desi_sim}
\end{figure*}

To see how resolution effects affected our bar fractions, we examined a low redshift ($z < 0.05$) sample of strongly and weakly barred galaxies identified using the automated classifications of GZ DESI \citep{walmsley_2023} for which we have bar length measurements available. This sample and the associated bar lengths have previously been used in \citet{geron_2023} and \citet{geron_2024}. The simulations of \citet{liang_2024} show that bars can be reliably detected as long as the ratio of the bar length ($L_{\rm bar}$) to \gls{FWHM} of the \gls{PSF} of the band is greater than 2. This threshold has previously also been proposed by other studies \citep{erwin_2018, xu_2024}. We can calculate what the angular size of all the bars in this new sample would be if their hosts would be at different redshifts. We then divide these angular sizes by the \gls{FWHM} of the DESI $r$-band, the JWST F444W band, and the JWST F200W band. The \gls{FWHM} of the DESI $r$-band is 1.18 arcsec \citep{dey_2019}. The \gls{FWHM} for the F444W band is 0.14 arcsec and is 0.064 arcsec for the F200W band\footnote{\url{https://jwst-docs.stsci.edu/jwst-near-infrared-camera/nircam-performance/nircam-point-spread-functions}}. We then note what fraction of bars in this sample have $L_{\rm bar}$ / FWHM $\geq$ 2 for each of these three bands. The results are shown in Figure \ref{fig:desi_sim}.

The leftmost panel shows the detection fraction of the sample when using the \gls{DESI} $r$-band. It very quickly drops off to 0, even for the strongest and longest bars. This clearly explains why studying the high redshift bar fraction with \gls{DESI} may not be very effective. In contrast, the detection fraction for the JWST F444W band, which has the highest \gls{FWHM} of all \gls{CEERS} bands, stays $>$0.7, even for weak bars. The detection fraction for the shorter wavelength F200W band is $\sim$99\%. This suggests that resolution is not the limiting factor for detecting bars with JWST.

Note that these simple simulations assume that there is no evolution in the length and strength of bars with redshift. This is not true, as bars are known to grow longer and stronger over time \citep{sellwood_1981,athanassoula_2003,martinez_valpuesta_2006,athanassoula_2013,algorry_2017}. This means that bars will be smaller at higher redshifts, which would make them more difficult to detect than what is assumed here. These simplified simulations serve as a useful initial approximation, but they highlight the need for more advanced approaches to estimate the bar detection fraction.


\bibliography{bibtex}{}
\bibliographystyle{aasjournal}



\end{document}
